\documentclass[referee]{raa}            % referee version: for submission
\usepackage{graphicx,times}             %for PS/EPS graphics inclusion, new
\usepackage{natbib}
\usepackage{amssymb,amsmath}
\bibpunct{(}{)}{;}{a}{}{,}

\usepackage[a4paper=true,dvipdfm=true,pagebackref=true]{hyperref}
\hypersetup{colorlinks = true, linkcolor = green, anchorcolor = red, citecolor = blue, filecolor = red, pagecolor = red, urlcolor = red}
\def\hi{\textsc{Hi~}}

\def\LaTeX{L\kern-.36em\raise.3ex\hbox{a}\kern-.15em
    T\kern-.1667em\lower.7ex\hbox{E}\kern-.125emX}

\numberwithin{equation}{section}

\usepackage{tabularx, booktabs, threeparttable, caption}

\newcolumntype{L}[1]{>{\raggedright\arraybackslash}p{#1}}
\newcolumntype{R}[1]{>{\raggedleft\arraybackslash}p{#1}}

\def\apjl{Astrophysical Journal Letters}
\def\aap{Astronomy and Astrophysics}

\def\apjs{The Astrophysical Journal Supplement Series}

\begin{document}

   \title{\textbf Forecasts of cosmological constraints from HI intensity mapping with FAST, BINGO \& SKA-I
%\,$^*$
%\footnotetext{$*$ Supported by the National Natural Science Foundation of China.}
}
%   \subtitle{I. Place Your Subtitle Here}

   \volnopage{Vol.0 (20xx) No.0, 000--000}      %%preserved for Editor. DOn't remove!
   \setcounter{page}{1}          %%starting page, preserved for Editor. DOn't remove!

   \author{Elimboto Yohana
      \inst{1,2,3,4}
   \and Yi-Chao Li
      \inst{3,4}
   \and Yin-Zhe Ma
      \inst{3,4}
   }
%% Here is an example of three authors come from different institutes.
%% For single author or all the authors from an institute, use "\inst{}" only

   \institute{Astrophysics and Cosmology Research Unit, School of Mathematics, Statistics $\&$ Computer Science, University of KwaZulu-Natal, Westville Campus, Private Bag X54001, Durban, 4000, 
   South Africa; {\it ey@tssfl.com}\\
%% Please give the E-mail address of the author, to whom future correspondence and
%% offprint requests will be sent.
        \and
             Dar Es Salaam University College of Education, A Constituent College of the University of Dar Es Salaam, P.O. Box 2329 Dar Es Salaam, Tanzania\\
        \and
             Astrophysics and Cosmology Research Unit, School of Chemistry and Physics, University of KwaZulu-Natal, Westville Campus, Private Bag X54001, Durban, 4000, South Africa\\
        \and 
            NAOC-UKZN Computational Astrophysics Centre (NUCAC), University of KwaZulu-Natal, Durban, 4000, South Africa\\
\vs\no
   {\small Received~~20xx month day; accepted~~20xx~~month day}}

\abstract{We forecast the cosmological constraints of the neutral hydrogen (HI) intensity mapping (IM) technique 
with radio telescopes by assuming 1-year of observational time. 
The current and future radio telescopes we consider here are FAST (Five hundred meter Aperture Spherical 
Telescope), BINGO (Baryon acoustic oscillations In Neutral Gas Observations), and 
SKA-I (Square Kilometre Array phase I) single-dish experiment. We also forecast the combined 
constraints of the three radio telescopes with {\it Planck}. We find that, the $1 \sigma$ errors of $(w_{0}, w_{a})$ for BINGO, FAST and SKA-I with respect
to the fiducial values are respectively, $(0.9293, 3.5792), (0.4083, 1.5878), (0.3158, 0.4622)$. This is equivalent to $(56.04\%, 55.64\%)$ and $(66.02\%, 87.09\%)$ 
improvements in constraining $(w_{0}, w_{a})$ for FAST and SKA-I relative to BINGO. 
Simulations further show that SKA-I will put more stringent constraints than both 
FAST and BINGO when each of the experiment is combined with \textit{Planck} 
measurement. The $1 \sigma$ errors for $(w_{0}, w_{a})$, BINGO + \textit{Planck}, 
FAST + \textit{Planck} and SKA-I + \textit{Planck} covariance matrices are respectively, 
$(0.0832, 0.3520), (0.0791, 0.3313), (0.0678, 0.2679)$, implying $(w_{0}, w_{a})$ 
constraints improvement of $(4.93\%, 5.88\%)$  for FAST + \textit{Planck} relative to 
BINGO + \textit{Planck} and an improvement of $(18.51\%, 23.89\%)$ in constraining 
$(w_{0}, w_{a})$ for SKA-I + \textit{Planck} relative to BINGO + \textit{Planck}. 
We also compared the performance of {\it Planck} data plus each single-dish experiment relative to {\it Planck} alone, and find that the reduction in $(w_{0}, w_{a})$ $1\sigma$ errors for each experiment 
plus {\it Planck}, respectively, imply the $(w_{0}, w_{a})$
constraints improvement of $(22.96\%, 8.45\%), (26.76\%, 13.84\%)$ and $(37.22\%, 30.33\%)$ for BINGO + {\it Planck}, FAST + {\it Planck} and SKA-I + {\it Planck} relative to {\it Planck} alone.
For the $9$ cosmological parameters in consideration, we find that, there is a trade-off
between SKA-I and FAST in constraining cosmological parameters, with each experiment being
more superior in constraining a particular set of parameters.
}

\authorrunning{E. Yohana, Y.-C. Li \& Y.-Z. Ma }            %author_head in even pages
   \titlerunning{Forecasts of \hi IM for FAST, BINGO $\&$ SKA-I}  % title_head in odd pages

   \maketitle
%% The author head (on even pages) and the title head (on odd pages) will be
%% automatically extracted from \author{} and \title{}. Whenever the title is too long,
%% you will be asked to supply a shorter one by inserting either \authorrunning{} or
%% \titlerunning{} before \maketitle. Anyway, you can specify your own heads.
%%
%%
%% Note: In the following text body of your manuscript, please note several differences from
%%       other major journals:
%% (1) \subsection{Please Capitalize the First Letter of Each Notional Word in Subsection Title}
%% (2) Please Capitalize the First Letter of Each Notional Word in all tables' captions

%
%________________________________________________ sections below
%
\section{Introduction}           %% first-level sections will be auto-capitalized
\label{section_6dfgs_intro}
Cosmology has unveiled our understanding of the universe that has increasingly matured over the last few decades. Up to this time, the study of the Universe has mostly given us a basic picture of how the 
Universe evolved and formed the large-scale structure. Many experiments dedicated to studying Universe throughout its entire history at various
epochs have been conducted so far, others are ongoing or planned to take off in the near future. Some of the notable surveys targeting the large-scale structures (LSS) of the Universe include a number of
galaxy redshift surveys such as 
the Two-degree-Field Galaxy Redshift Survey (2dFGRS, \cite{2dF}), 
the WiggleZ Dark Energy Survey \citep{WiggleZ1, WiggleZ2, 2014MNRAS.441.3524K},
the Six-degree-Field Galaxy Survey (6dFGS, \citep{6dFGS, 6dFcorr}), and
the Baryon Oscillation Spectroscopic Survey (BOSS, \cite{SDSS-BOSS}), which 
is the third stage of the Sloan Digital Sky Survey (SDSS, \cite{SDSS, SDSS-III, SDSS-III_2017}).
Recently, the Dark Energy Survey (DES), \cite{DES_survey}
reported their cosmological constraints with the $1$-year data~\citep{DES1y, DES1y2}.
Future optical surveys that aim to use larger and sensitive 
telescopes at a variety of high redshifts, such as DESI \citep{DESI, DESI_2016},
LSST \citep{LSST, 2009arXiv0912.0201L, 2018arXiv180901669T}, {\it Euclid} \citep{Euclid} and {\it WFIRST} \citep{WFIRST},
have been proposed and some of the constructions are under-way.
Up-to-date, galaxy-redshift surveys have made significant progress in surveying the large-scale 
structures of the Universe. In order to do precision cosmology, one would need to detect sufficiently large samples of \hi-emitting galaxies. This is a huge task, since at higher redshifts the galaxies 
look essentially very faint \citep{Late-time-cosmos,Line-IM,21cm-review}. 

In radio astronomy, the observation of $21$cm spectrum line emitted by the
neutral hydrogen (\hi) in the deep space
provides a rich tool for understanding the cosmic evolution. After the cosmic
reionization, the hydrogen outside the galaxies is ionized. But massive amount of \hi shielded from ionizing UV photons,
resided within the dense gas clouds embedded in galaxies, as these gas clouds cooled and collapsed to form stars.
As a result, the quantity and distribution of \hi is related to the evolution 
of galaxies and the cosmic surveys in the radio band, whose origin and evolution is highly related to the structure formation history, and the nature of cosmic expansion~\citep{21cm-review, Line-IM}. 

At lower redshift $z \lesssim 0.1$, \hi can be detected with the $21$cm emission and absorption lines
from its hyperfine splitting. At redshift greater than $2.2$, \hi can also be
detected via optical observation of the Ly$\alpha$ absorption line against
the background bright sources \citep{HIPASS2, ALFALFA2}. 
At intermediate redshift, the $21$cm emission
line of each individual galaxies is too faint to be detected. However, instead 
of cataloging individual galaxies, the intensity mapping (IM) method measures the total \hi flux from many galaxies, and can be used for 
LSS studies \citep{IM-Chang, IM-Loeb}. With the \hi IM method, 
\cite{2010Natur.466..463C} firstly reported the measurements of cross-correlation 
function between the \hi map, observed with Green Bank Telescope (GBT), and DEEP2 optical 
redshift survey \citep{2001defi.conf..241D}. With the extended GBT \hi survey and 
WiggleZ Dark Energy Survey, the cross-power spectrum between
\hi and optical galaxy survey was also detected
\citep{GBT}. Recently, another \hi survey with Parkes telescope
reported the measurements of cross-power spectrum with 2dF optical galaxy survey
\citep{2017arXiv171000424A}. So far, the auto-power spectrum of \hi IM survey
is still not detected \citep{2013MNRAS.434L..46S}, because of the
contamination of the foreground residuals.

There is a number of current and future experiments targeting \hi IM.
These experiments increasingly comprise of 
wide-field and sensitive radio telescopes or interferometers, 
such as BAOs in Integrated Neutral Gas Observations (BINGO, \cite{2014arXiv1405.7936D}), 
Canadian Hydrogen Intensity Mapping Experiment (CHIME, \cite{CHIME}), 
Tianlai \citep{Tianlai} and
Hydrogen Intensity and Real-time Analysis eXperiment (HIRAX, \cite{HIRAX}).
Besides of the special designed telescopes or interferometers, several larger
single-dish telescopes and interferometers, such as 
the Five-hundred-meter Aperture Spherical radio Telescope (FAST, \cite{Nan}), 
Square Kilometer Array (SKA, \cite{SKAI, SKAI-2, SKA-science}) or MeerKAT \citep{MeerKLASS}, 
are also planned for \hi IM survey.

This paper aims to use \hi IM to forecast how the future \hi  experiments,
such as BINGO, FAST and SKA Phase I (SKA-I), will constrain various 
cosmological parameters. 

FAST is the world-largest single-dish telescope
for high resolving power. BINGO is a medium sized single-dish telescope with 
special design~\citep{BINGO-updates}. SKA-I is a telescope array in single-dish 
autocorrelated mode suitable for probing large volumes over very large cosmological scales. 
These experiments are the next-generation LSS surveys which can be used to learn and address excellent techniques of \hi IM surveys. Our aim is to simultaneously consider three experiments 
whose nature and designs categorically represent many future \hi IM probes. We will 
present quantitative and qualitative comparison between their future prospects, while addressing the range of
expected performances, limitations and challenges that may
accompany these experiments. 

We develop a forecast framework motivated and 
guided by physical experimental design and set-ups, correctly transformed 
into mathematics and computer simulations. We believe that, this clear scientifically 
motivated forecast study will substantially provide testable predictions and determine paths and feasibility 
of the future \hi IM experiments.

The outline of the paper is as follows: In Section \ref{section_2}, we briefly describe the three future experiments, namely, BINGO, FAST and SKA-I.
In Section \ref{section_3}, we discuss and summarize the mathematical derivation of the tomographic angular power spectrum and introduce the thermal noise power spectrum
as residuals of various contaminants after applying foreground removal techniques. This spectrum of noise is related
to various observable experimental parameters. We further show the calculation of noise power spectrum in Section \ref{noise_subsection} and tomographic power spectrum to compute Fisher matrix 
in Section \ref{fisher_matrix_subsec}. Noise power spectrum 
together with tomographic angular power spectrum are prime tools for computing Fisher matrix via the maximum likelihood estimation (MLE)~\citep{Modern-cosmos, coaxing-21cm}.
We present forecasts of cosmological constraints in Section \ref{section_4} based on various cosmological parameters of our choice and analyze the results. In this Section, we also define the cosmological parameters used 
and present various FAST, BINGO
and SKA-I experimental parameter specifications. We summarize our forecasts in Section \ref{section_5}, and finally, we conclude our paper in Section \ref{section_6}. 

Unless otherwise stated, we adopt a spatially-flat $\Lambda$CDM cosmology model with fiducial parameters listed in Table~\ref{cosmological_parms_description} \citep{Planck-collaboration, Planck-collaboration1}.

%% Authors can give a citation as 'Michel et al. 1992'.
%% You may also use \cite, \citep and \citet for citation, and use Table~1 or Figure~1
%% and so forth. Using \ref and \label for cross-references of Tables/Figures
%% is a good way in adjusting/adding/removing text, tables or figures.
\begin{table*} 
\begin{center}
\caption{The cosmological parameters in our study as the best-fitting parameters in ~\cite{Planck-collaboration1}.}
\label{cosmological_parms_description}
\begin{tabular}{@{}cc|>{\raggedright\arraybackslash}m{100mm}}\hline
                       Parameter & Fiducial value & \hspace{40mm} Description\\\hline 
$\Omega_{\rm b}h^{2}$ & $0.0226$ & Fractional baryon density today \\
$\Omega_{\rm c}h^{2}$ & $0.112$ & Fractional cold dark matter density today \\
$w_{0}$               & $-1.00$ & Dark energy equation of state, from the relationship $w(a) = w_{0} + (1-a)w_{a}$ \\
$w_{a}$               & $0.00$ & Dark energy equation of state, from the relationship $w(a) = w_{0} + (1-a)w_{a}$ \\
${\rm ln}(10^{10}A_{{\rm s}})$ & $3.089$ & Log power of the primordial curvature perturbations, ($k_{0} = 0.05 \ \text{Mpc}^{-1}$) \\
$H_{0}$                        & $70.00$ & The Hubble constant (current expansion rate in $\text{km} \ \text{s}^{-1} \ \text{Mpc}^{-1}$) \\
$N_{\rm eff}$                  & $3.046$ & Effective number of neutrino-like relativistic degrees of freedom\\
$n_{{\rm s}}$                  & $0.96$  & Scalar spectrum power-law index ($k_{0} = 0.05 \ \text{Mpc}^{-1}$)  \\
$\Sigma m_{\nu}/94.07 \ \text{eV}$ & $0.00064$ & The sum of neutrino masses in eV \\\hline\hline
\end{tabular}
\end{center} 
\end{table*} 

\section{Intensity mapping projects}
\label{section_2}
BINGO, FAST and SKA-I experiments are potentially suitable for surveying \hi intensity maps of the 
Universe and open avenues for doing a wide range of sciences. 
In this section, we briefly describe each of these three future experiments for studying the IM of neutral hydrogen.

\subsection{BINGO}
BINGO project is proposed to be built in Brazil and aims to map \hi emission at 
redshift range $0.13-0.48$ ($960\,{\rm MHz}\sim 1260\,{\rm MHz}$). BINGO will map approximately
$15{}^\circ$ strip of the sky to measure the \hi power 
spectrum and detect for the first time, BAOs at radio frequencies. BINGO expected
design is a dual-mirror compact 
antenna test range telescope with a $40$ m primary mirror and an offset focus,
proposed to have receiver array containing between 50 - 60 feed horns, with a 
$90\, {\rm m}$ focal length. 
For more details about BINGO construction and its prospective capabilities, 
please refer to~\cite{BINGO-IM, BINGO, BINGO-updates}. 

\subsection{FAST}
\label{fast_subsection}
FAST is a ground-based radio telescope built within a Karst depression in 
Guizhou province of Southwest China. 
The L-band receiver is build with $19$ beams and the multibeam receiver will
increase the survey speed \citep{Nan}. FAST is believed to be the most sensitive single dish telescope, 
covering a wide frequency range from $70\, \text{MHz} - 3\, \text{GHz}$
and potentially large area of up to
$25,000\, \text{deg}^{2}$. 
Here we consider a survey area of $10,000\,\text{deg}^{2}$, approximately equivalent to the one 
used by \cite{SDSS-DR11-DR12}. A chosen survey area reasonably suffices our current study, and is moderate by taking 
into consideration 
other experimental parameters and design factors.
In addition, this choice is also 
potentially suitable for any future FAST-SDSS cross-correlation studies. 
For the \hi IM survey with FAST, we consider a frequency range of 
$950\,{\rm MHz}\sim 1350\,{\rm MHz}$. FAST has a diameter of 500 meters, but the 
illuminated aperture is 300 meters. For full details of FAST engineering and its
capabilities please refer to \cite{Nan, 21-cm_IM_with_FAST}.

\subsection{SKA-I}
SKA project, currently under development, 
is basically an interferometry array. The project is a two-stage development, comprising of SKA Phase I and SKA Phase II \citep{SKAI, SKAI-2, SKA-science}. The first stage (SKA Phase I) radio astronomy 
facility is split and shared between South Africa (SKAI-MID) -- hosted in 
Karoo Desert, and an aperture array in Australia,  SKA-LOW Phase I (SKAI-LOW). SKAI-MID plans to build $133$, $15 \ {\rm m}$ diameter dishes and will
incorporate $64$ dishes MeerKAT array \citep{MeerKLASS, MeerKAT_DES_2017} each with $13.5 \ {\rm m}$ diameter, that have already been constructed in the Karoo Desert. Note that, SKA-I telescope specifications 
used for our study have been subject to changes as the project go through various levels of revision \citep{Bull_2016}, see recent updates \cite{SKAI_2018_Red_Book}.
Due to the weak resolution requirement for \hi IM, we ignore the
cross correlation between dishes, which means the SKAI-MID array is working as $133$
single dishes, with an extension of $64$, $13.5 \ {\rm m}$ MeerKAT array dishes. 
We therefore consider tentative experimentation with SKAI-MID Band 1 (excluding MeerKAT array), hereafter referred to as SKA-I, at frequencies $350\,{\rm MHz}\sim 1050\,{\rm MHz}$
for the full $133$ antennae for a total survey area of $10,000 \ \text{deg}^{2}$. We however make the same choice of survey area as for 
FAST for similar reasons as explained in Subsection \ref{fast_subsection}. For full details about BINGO, 
FAST and SKA-I experimental design, see Table~\ref{experimental_parameters} .

\begin{table} 
\begin{center}
\caption{The experiment parameters for FAST, BINGO and SKA-I. 
$D_{\mathrm{dish}}$ is the illuminated aperture \citep{Yi-Chao}.}
\label{experimental_parameters}
\begin{tabular}{@{}c|ccc}\hline\hline 
                                         & FAST        & SKA-I       & BINGO\\\hline
$\nu_{\mathrm{min}}[\mathrm{MHz}]$      & $950$      & $350$       & $960$\\
$\nu_{\mathrm{max}}[\mathrm{MHz}]$      & $1350$      & $1050$      & $1260$\\
$\Delta\nu[\mathrm{MHz}]$             & $10$        & $10$        & $10$\\
$n_{\nu} (n_z)$                         & $40 $       & $70$        & $30$\\
$D_{\mathrm{dish}}[\mathrm{m}]$         & $300$       & $15$        & $25$\\
$N_{\mathrm{ant}}\times N_{\mathrm{feed}}$& $1\times19$ & $133\times1$& $1\times60$\\
$t_{\mathrm{TOT}}[\mathrm{yr}]$         & $1$         & $1$         & $1$\\
$T_{\mathrm{rec}}[\mathrm{K}]$    & $25$        &  ${\rm Eq.~(\ref{SKAI_T_rec})}$  & $50$\\
$S_{\mathrm{survey}}[\mathrm{deg}^2]$   & $10,000$    & $10,000$    & $3,000$\\\hline\hline
\end{tabular}
\end{center} 
\end{table} 

\section{Method}
\label{section_3}
\subsection{Tomographic Angular Power Spectrum}
In our forecast, we consider the tomographic angular power spectrum of \hi for the 
$i$-th and $j$-th redshift bins given by
\begin{equation}\label{tomography}
    C_{\ell}^{ij} = 4\pi T_{b}^{ij}\int {\rm d} \ln k \Delta^{2}(k) \Delta^{W}_{T_{\rm b},\ell}(k)\Delta^{W'}_{T_{\rm b},\ell}(k), 
\end{equation}
where $\Delta^{2}(k)$ is the dimensionless power spectrum of primordial curvature 
perturbation. Here, $T_{\rm b}^{ij} = T_{\rm b}(z_i)T_{\rm b}(z_j)$, is the multiplication
of \hi mean brightness temperature \citep{IM-Chang} of the $i$-th and $j$-th 
redshift bins, with 
\begin{equation}
T_{\text{b}}(z) = 0.39 \ {\rm mK}\left( \frac{\Omega_{\rm \hi}}{10^{-3}} \right)\left( \frac{1+z}{2.5} \right)^{0.5} \left( \frac{\Omega_{\rm m} + (1+z)^{-3}\Omega_{\Lambda}}{0.29} \right)^{-0.5}
\end{equation}
where $\Omega_{\text{\hi}}$ is the fractional of \hi density assumed to be $0.62 \times 10^{-3}$ \citep{2013MNRAS.434L..46S} and $\Delta^{W}_{T_{\rm b},\ell}(k) \equiv \Delta^{W}_{T_{\rm b},\ell}(\mathbf{h})/\mathcal{R}(\mathbf{k})$~\citep{Hall13}. The transfer function is
\begin{equation}\label{transfer_function} 
%%W_{\ell}(k) = \int {\rm d} \chi \frac{{\rm d}N_{g}(\chi)}{{\rm d}\chi}j_{\ell}(k\chi)b_{\text{\hi}}(\chi(z),k)T_{\delta}(\chi, k),
\Delta^{W}_{T_{\rm b},\ell}(\mathbf{k})=\int^{\infty}_{0}{\rm d}z W(z) \Delta_{T_{\rm b},\ell}(\mathbf{k},z),
\end{equation}
which is an integration of the temperature fluctuation over the band-width $W(z)$. The temperature fluctuation, for each $\ell$ (projected mode) for each wavenumber $\mathbf{k}$ and redshift bin $z$ is
\begin{eqnarray}
\Delta_{T_{\rm b},\ell}(\mathbf{k},z)&=& \delta_{n}j_{\ell}(k\chi)+\frac{kv}{\mathcal{H}}j''_{\ell}(k\chi)+\left( \frac{1}{\mathcal{H}}\dot{\Phi} + \Psi\right)j_{\ell}(k\chi) \nonumber \\
&-& 
\left(\frac{1}{\mathcal{H}}\frac{{\rm d} \ln(a^{3}\bar{n}_{\hi})}{{\rm d} \eta}-\frac{\dot{\mathcal{H}}}{\mathcal{H}^{2}}-2 \right)\left[
\Psi j_{\ell}(k\chi)+vj'_{\ell}(k\chi)+\int^{\chi}_{0}\left(\dot{\Psi}+\dot{\Phi} \right)j_{\ell}(k \chi'){\rm d}\chi' \right],\nonumber \\
\end{eqnarray}
where $j_{\ell}$ is the spherical Bessel function, $\delta_{n}$ is the $\hi$ density contrast, and the second term $kvj''_{\ell}(k\chi)/\mathcal{H}$ is the redshift space distortion term~\citep{Hall13}.

Here we work in the tomographic power spectrum in $\ell$-space of multiple redshift (frequency) slices. We notice that there are several previous works which implemented the forecasts in 
3-D $k$-space~\citep{SKAI,Late-time-cosmos}. There are some advantages that the tomographic 2-D power spectrum in $\ell-$space has compared to 3-D power spectrum in $k$-space. 
The 3-D power spectrum in $k$-space has the following disadvantages:

\begin{itemize}
 \item It assumes plane-parallel, so it cannot encompass wide angle correlations;
\item it cannot include lensing effect, either;
\item in the analysis of 3-dimensional power spectrum, the redshift bins are typically wide, this neglects the evolution of background within bins; and
\item it requires a fiducial model which must be assumed to relate redshift to distance. 
\end{itemize}
In addition, the tomographic angular power spectrum can easily be applied to perform cross-correlations between 21cm images and other large-scale structure tracers at the same redshift. Due to 
these reasons, our approach thus has some advantages over 3-dimensional power spectrum, and we find it worthy investigating as we have done so in this work. Full details regarding 
the advantages of using the tomographic angular power spectrum are found in \cite{Shaw_2008, Di_Dio_2014, Tansella_2018, Camera_2018}.

\subsection{Noise}
\label{noise_subsection}
Noise for the single-dish intensity mapping experiment is given by
\begin{equation}
N_{\ell}^{ij} = \delta^{ij}N_{\ell}^{\text{\hi}} = \frac{\delta^{ij}T_{\text{sys}}^{2}S_{\text{survey}}}{(N_{\text{ant}}N_{\text{feed}}t_{\text{TOT}}\Delta \nu)},
\end{equation}
where $N_{\text{ant}}$ is the number of antennae, $N_{\text{feed}}$ is the total number of feed horns and $t_{\text{TOT}}$ is the total observational time.

BINGO and FAST system temperatures are given by
\begin{equation}
T_{\text{sys}} = T_{\text{rec}} + T_{\text{sky}},
\end{equation}
whereas SKA-I system temperature is modeled by adding ground spill-over~\citep{SKAI_2018_Red_Book}
\begin{equation}\label{SKAI_T_sys}
 T_{\rm sys} = T_{\rm rec} + T_{\rm spl} + T_{\rm sky}.
\end{equation}
Here, $T_{\rm spl} \approx 3 \ {\rm K}$ is the spill-over contribution.

Furthermore, $T_{\text{rec}}$ is the receiver temperature particular to each telescope model. BINGO and FAST receiver temperatures are
presented in Table ~\ref{experimental_parameters}, where for SKA-I
\begin{equation}\label{SKAI_T_rec}
 T_{\rm rec} = 15 \ {\rm K} + 30 \ {\rm K} (\nu ({\rm GHz}) - 0.75)^{2}. 
\end{equation}

Basically, all three telescopes see the same sky, so we model their sky temperature contribution as 

 \begin{equation}\label{SKAI_T_gal}
 T_{\rm sky} = T_{\rm gal} +  T_{\rm CMB},      
\end{equation}
with 
\begin{equation}
 T_{\rm gal} \approx 25 \ {\rm K} (408 \ {\rm MHz}/\nu)^{2.75}
\end{equation}
being the contribution from our Milky Way Galaxy for a given frequency $\nu$, and  $T_{\rm CMB} \approx 2.73 \ {\rm K}$ the CMB temperature. One can refer to \cite{SKAI_2018_Red_Book} for more information regarding system 
temperature, and Table~\ref{experimental_parameters} for the detailed list of exact values of each experimental parameters considered in our forecast.

Genuinely speaking, 21cm intensity maps highly suffer from contaminations due to foregrounds, such as Galactic synchrotron emission, extragalactic point sources, and atmospheric noises. Thus, application of 
foregrounds cleaning techniques
are inevitably important in order to mitigate these contaminations. However, there is always some level of contamination residuals after applying such techniques. Therefore, the cross-correlation of noises 
between different frequency bins may not 
completely be negligible. So  the elements of the noise matrix $N_{\ell}$ have been treated under some simplified assumptions~\citep{Yi-Chao}. 

\subsection{Fisher Matrix}
\label{fisher_matrix_subsec}
\begin{figure}
\begin{center}
\includegraphics[width=8.8cm]{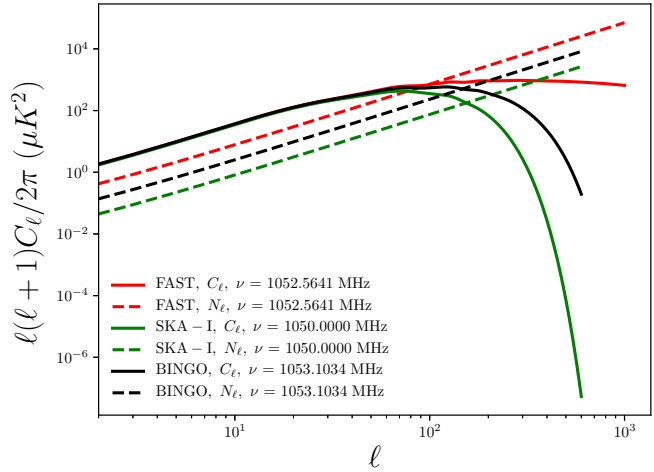} 
\caption{The noise power spectra $N_{\ell}$ (dashed line) and beam convolved angular power spectra, $C_{\ell}$ 
(solid line) for FAST (red), BINGO (black) and SKA-I (green) at approximately
overlapped frequencies. As expected, we see that the angular power spectra 
have almost the same profile at large scales but deviating with increase 
in number of multipoles, $\ell$. Beyond $\ell = 150$, angular power spectra
for BINGO and SKA-I more rapidly become insignificant than noise compared to 
FAST angular power spectrum.}
 \label{overlap_frequency}
 \end{center}
\end{figure}

We perform the Fisher matrix analysis to explore the potential of the future 
\hi IM experiments for constraining the cosmological parameters.
Assuming that the maximum likelihood estimation can be well approximated by multivariate
Gaussian function, the Fisher matrix $\mathcal{F}$ is then a good approximation 
of the inverse of the parameter covariance. The Fisher matrix is expressed as,
\begin{equation}\label{fisher_matrix}
\mathcal{F}_{\alpha \beta} = f_{\text{sky}}\sum_{\ell_{\text{min}}}^{\ell_{\text{max}}}\left(\frac{2\ell + 1}{2}\right)\text{tr}\left[C_{\ell, \alpha}\Sigma_{\ell}C_{\ell, \beta}\Sigma_{\ell}\right],
\end{equation}
in which, the total noise inverse matrix is given by 
\begin{equation}
\Sigma_{\ell} = (C_{\ell} + N_{\ell})^{-1}.
\end{equation}
Here, $N_{\ell}$, the noise power spectrum, is an $n_{\nu} \times n_{\nu}$ matrix. 
We assume that, noises between $i$-th and $j$-th frequency channels ($i \neq j$) are uncorrelated, and thus $N_{\ell}$ is a diagonal matrix. 
The tomographic angular power spectrum, $C_{\ell}$, is an $n_{\nu} \times n_{\nu}$
matrix, and each element of $C_{\ell}$ is the \hi cross angular power spectrum 
of the $i$-th and $j$-th redshift bins. 
Furthermore, we multiply the $C_{\ell}$ with the window function 
for $i,j$-th frequency channels, 
\begin{equation}\label{window_function}
    W_{\ell, ij} = {\rm e}^{-{\ell}^2\left(\sigma_i^2 + \sigma_j^2\right)/2},
\end{equation}
which is simply the multiplication of the Fourier space Gaussian beam function
at the $i$-th and $j$-th frequency channels. 
In this case, 
\begin{equation}
    \sigma_i = \theta_{\rm FWHM}/ (\sqrt{8\ln(2)}) {\simeq} \ 0.4245\theta_{\rm FWHM},
\end{equation}
where $\theta_{\rm FWHM}=1.22\lambda/D_{\rm dish}$ is the full width at half-maximum of the beam. 
The window function (\ref{window_function}) implies that, 
at large values of $\ell$, corresponding to small angular scales, 
it falls off rapidly as depicted by \hi angular power spectra in Figure~\ref{overlap_frequency}.

For all cosmological constraints, we ignore the monopole and dipole moments, and consider a multipole moments 
range from $\ell = 2$ to $\ell = 600$ for forecast with BINGO and SKA-I, and $\ell = 2$ to $\ell = 1000$ 
for FAST. This range of $\ell$ is chosen for each telescope to make sure within each range of $\ell$ the signal-to-noise ratio is significant, which contributes to the constraints of cosmological parameters. 
For very high $\ell$, the beam function makes the signal to be below the noise power spectrum, so adding high-$\ell$ of the power spectrum does not improve the constraints.

Here we vary $9$ parameters which are shown in Table~\ref{cosmological_parms_description}. Therefore our Fisher matrix (Eq.~(\ref{fisher_matrix})) is a $9 \times 9$ matrix. To see how we can tighten up 
constraints with {\it Planck} satellite data, we used the best-fit $\Lambda$CDM CMB power spectra from the baseline {\it Planck} chains, that include {\it TT + TE + EE + lensing}, taken from the Planck 
Legacy Archive website -- Cosmology section \footnote{\url{https://pla.esac.esa.int/\#cosmology}} to compute the {\it Planck} covariance. We then make an entry-wise addition of the 
{\it Planck} Fisher matrix (the inverse of the {\it Planck} covariance \citep{Planck-collaboration1}) parameter respectively, 
to the corresponding parameter entries in the resulting 
Fisher matrix calculated using the formulae (\ref{fisher_matrix}) for each particular \hi IM experiment. The model cosmological parameters whose values were varied by making entry-wise 
addition of the \textit{Planck} Fisher matrix correspondingly, to BINGO, FAST and 
SKA-I Fisher matrices are $\Omega_{\rm b}h^{2}, \ \Omega_{\rm c}h^{2}, \ w_{0}, \ w_{a}, \ {\rm ln}(10^{10}A_{{\rm s}}), \ H_{0}$ 
and $n_{{\rm s}}$. The rest of \hi experiment parameters, namely, $N_{\rm eff}$ 
and $\Sigma m_{\nu}/94.07 \ \text{eV}$, were omitted in order to only consider parameters that conform with the \textit{Planck} chains dataset, \cite{Planck-collaboration1} (believed to set strongest 
constraints to cosmological parameters), that we used to compute the \textit{Planck} Fisher matrix under consideration. 
\section{Results and Discussion}
\label{section_4}

In this section we present two sets of forecast results, 
the first one detailing various cosmological constraints comparison between 
FAST, BINGO and SKA-I, and the
second one showing relative constraining capabilities by combining each of the 
three experiments with \textit{Planck}. \textit{Planck} covariance matrix includes TT + TE + EE + lensing, but 
throughout this paper,  we will use a shorthand \textit{Planck} to mean \textit{Planck} + TT + TE + EE + lensing.
Table~\ref{stds_FAST_BINGO_SKAI} shows the $1 \sigma$ errors for the marginalized parameter constraints for each of
these experiments. In our simulations, for all the three experiments, 
we fix the frequency bandwidth to be $10$ MHz, unless stated otherwise.
More specifically, the frequency (or equivalently redshift) division is done with uniformly spacing of channels, each of width $10$ MHz  or $1$ MHz, depending on the tests performed. 
Which means for standard tests carried with a channel width of $10$ MHz, we used $30$ redshift/frequency bins for BINGO, $40$ redshift bins for FAST and 70 redshift bins for SKA-I, while 
for tests carried with $1$ MHz channel width, we used $300$, $400$ and $700$ redshift bins, respectively, for BINGO, FAST and SKA-I. We have used significantly narrower channel width 
compared to most of the previous works, for the benefits we have motivated in the later sections.  Roughly, the central value of each channel width was used in calculations, that's 
the sum of the lower and upper margins divided by $2$. The central value of the bin is a good approximation for sufficiently narrower bins in that we can neglect evolution of 
cosmological functions/backgrounds within each redshift bin, because most of the relevant functions coupled in calculations of the angular power spectra vary slowly with 
redshift; instead, the evolving cosmological functions are fixed to their values at the central redshift of the bin, the choice which is however motivated by \cite{Late-time-cosmos}.
Full telescope specifications 
we used for simulations are presented in Table \ref{experimental_parameters},
and the descriptions of the cosmological parameters used in forecast are given in 
Table \ref{cosmological_parms_description}. 
We use {\tt Camb\_sources} \citep{2011PhRvD..84d3516C} to compute the raw tomographic 
angular power spectra Eq. (\ref{tomography}) and another code  we 
developed to simulate forecasts of cosmological parameter constraints
via Fisher matrix (Subsection \ref{fisher_matrix_subsec}). We will then compare the forecasted constraints between these different experiments.

\subsection{Dark Energy Constraints}
\label{dark_energy_constraints}
\begin{figure}
\begin{center}
 \includegraphics[width=8.8cm]{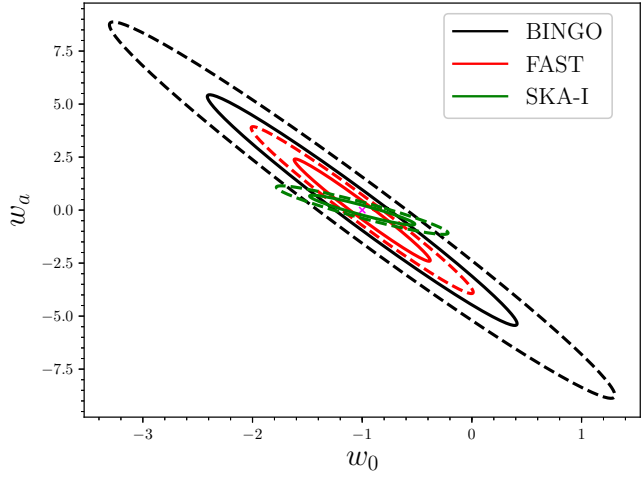} 
 \caption{$w_{0}$ versus $w_{a}$,  $1\sigma$ (solid line) and
 $2 \sigma$ (dashed line) cosmological constraints for 
 FAST (red), BINGO (black) and SKA-I (green).}
 \label{w0_wa_FAST_BINGO_SKAI}
 \end{center}
\end{figure}

We present two separate analysis, the first one is to show how FAST, BINGO, SKA-I can comparatively constrain the dark energy equation of state (EoS) in the form of  
$w(a) = w_0 + w_a(1-a)$ (``Chevallier-Polarski-Linder parametrization''~\citep{Chevallier01,Linder03}) and the second one is to show how each of these experiments, 
FAST, BINGO, SKA-I plus \textit{Planck} data
can constrain the dark energy equation of state. Figure~\ref{w0_wa_FAST_BINGO_SKAI} shows that FAST will
constrain the dark energy equation of state better than BINGO, and possibly 
than many other currently known single-dish \hi IM approach counterparts. But SKA-I 
puts more stringent constraints to the dark energy equation of state than both 
BINGO and FAST. The $1 \sigma$ errors from $(w_{0}, w_{a})$ 
covariance matrices for BINGO, FAST and SKA-I are respectively, $(0.9293, 3.5792), (0.4083, 1.5878), (0.3158, 0.4622)$.

To compare the relevant improvement on $1 \sigma$ errors from BINGO to FAST and SKA-I, we consider 
the largest error of the three experiments for a 
particular parameter and find out the fractions of the 
errors that have been reduced with respect to it. For $w_{0}$, 
the error is reduced by $(0.9293 - 0.4083)/0.9293 = 56.06\%$ and $(0.9293 - 0.3158)/0.9293 = 66.02\% $ for FAST and SKA-I with respect to BINGO. For $w_{a}$, it is 
$(3.5792 - 1.5878)/3.5792 = 55.64\%$ and $(3.5792 -  0.4622)/3.5792 = 87.09\% $ for FAST and SKA-I with respect to BINGO.
Therefore we can see quite significant improvement of FAST and SKA-I for future constraints on dark energy equation of state. Although the parameters $w_{0}$ and $w_{a}$
have some degeneracy, the joint constraints with {\it Planck} can improve significantly the constraints.

\begin{figure}
\begin{center}
 \includegraphics[width=8.8cm]{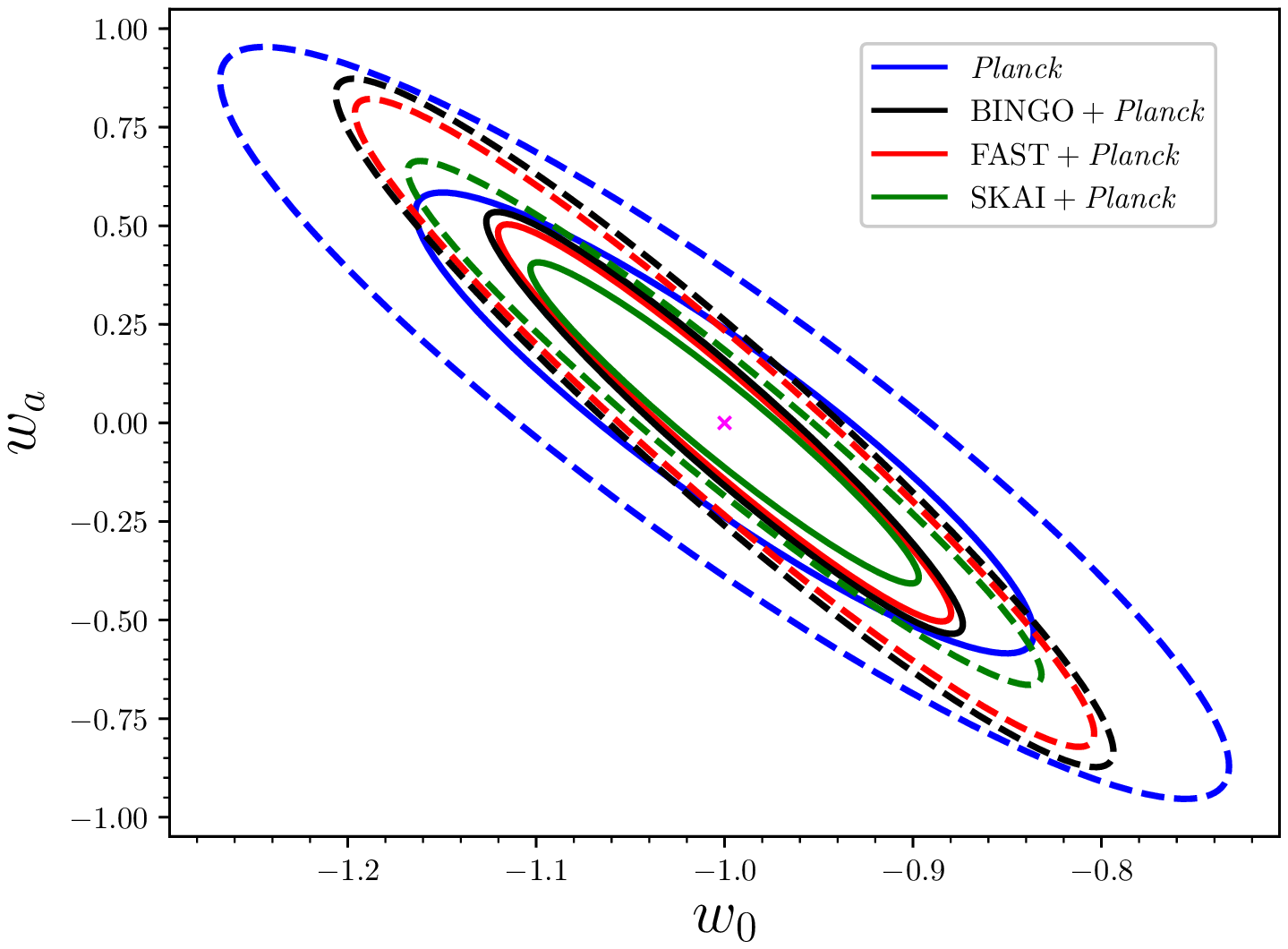} 
 \caption{$w_{0}$ versus $w_{a}$,  $1 \sigma$ (solid line) and $2 \sigma$ 
 (dashed line) cosmological constraints for {\it Planck} (blue), FAST + {\it Planck} (red), 
 BINGO + {\it Planck} (black) and SKA-I + {\it Planck} (green).}
 \label{w0_wa_FAST_BINGO_SKAI_plus_Planck}
 \end{center}
\end{figure}

The fact that FAST will do better than BINGO to constrain the dark energy
equation of state remains unchanged if each of the experiment is individually combined with \textit{Planck} data, as shown in Figure~\ref{w0_wa_FAST_BINGO_SKAI_plus_Planck}. This observation is valid and is 
however supported 
by \cite{FAST_IM}, although in their paper they have used a 
different set of experimental parameters. As previously observed from 
simulations in Figure~\ref{w0_wa_FAST_BINGO_SKAI}, again, Figure~\ref{w0_wa_FAST_BINGO_SKAI_plus_Planck} shows that SKA-I will put 
more stringent constraints than both FAST and BINGO when each of the 
experiment's Fisher matrix is added to \textit{Planck} Fisher matrix. The $1 \sigma$ errors for 
$(w_{0}, w_{a})$, BINGO + \textit{Planck}, FAST + \textit{Planck} and 
SKA-I + \textit{Planck} covariance matrices are respectively, $(0.0832, 0.3520), (0.0791, 0.3313), (0.0678, 0.2679)$, implying $(w_{0}, w_{a})$ 
constraints improve by $(4.93\%, 5.88\%)$ in error reduction for FAST + \textit{Planck} relative 
to BINGO + \textit{Planck} and an improvement of $(18.51\%, 23.89\%)$ error reduction in constraining $(w_{0}, w_{a})$
for SKA-I + \textit{Planck} relative to BINGO + \textit{Planck},
see Table \ref{stds_FAST_BINGO_SKAI}. It is very clear that, all three experiments improve dark energy constraints 
tremendously when the \textit{Planck} Fisher matrix is added to each of the respective experiment's Fisher matrix.

To benchmark the performance of each single-dish experiment combined with {\it Planck} relative to
{\it Planck} alone, we find that the $(w_{0}, w_{a})$ $1\sigma$ errors for {\it Planck}, BINGO + {\it Planck},
FAST + {\it Planck} and SKA-I + {\it Planck} are respectively, $(0.1080, 0.3845), (0.0832, 0.3520), (0.0791, 0.3313)$ and $(0.0678, 0.2679)$.
The reduction in $(w_{0}, w_{a})$ $1\sigma$ errors for each experiment plus {\it Planck}, respectively, imply the $(w_{0}, w_{a})$
constraints improvement of $(22.96\%, 8.45\%), (26.76\%, 13.84\%)$ and $(37.22\%, 30.33\%)$ for BINGO + {\it Planck}, 
FAST + {\it Planck} and SKA-I + {\it Planck} relative to {\it Planck} alone. Table \ref{stds_FAST_BINGO_SKAI}, Figure~\ref{FAST_BINGO_SKAI_plus_Planck}, and Figure~\ref{FBS_plus_Planck_rel_cons_improvement}  
summarize how the {\it Planck}-each-single-dish experiment joint constraints improve relative to the {\it Planck} data 
constraints alone for all the cosmological parameters considered.

\begin{table*} 
\begin{center}
\caption{$1 \sigma$ errors for FAST, BINGO, SKA-I and {\it Planck} covariance 
matrices, and those obtained from covariance matrices resulting 
from combination of each of the FAST, BINGO and SKA-I experiment's 
Fisher matrix with {\it Planck} Fisher matrix.}
\label{stds_FAST_BINGO_SKAI}
\begin{tabular}{@{}cccc|c|ccc} \hline\hline 
                                         & FAST   & BINGO  & SKA-I & {\it Planck} &FAST + {\it Planck} & BINGO + {\it Planck} & SKA-I + {\it Planck} \\\hline \hline
$\Omega_{\rm b}h^{2}$          &  $0.0090 $  &  $0.0168 $  & $0.0072 $ & $0.0002$ &    $0.0001 $  &  $0.0001 $  & $0.0001 $ \\
$\Omega_{\rm c}h^{2}$          &  $0.0061 $  &  $0.0133 $  & $0.0115 $ & $0.0014$ &    $0.0011 $  &  $0.0012 $  & $0.0008 $\\
$w_{0}$                        &  $0.4083 $  &  $0.9293 $  & $0.3158 $ & $0.1080$ &    $0.0791 $  &  $0.0832 $  & $0.0678 $ \\
$w_{a}$                        &  $1.5878 $  &  $3.5792 $  & $0.4622 $ & $0.3845$ &    $0.3313 $  &  $0.3520 $  & $0.2679 $\\
${\rm ln}(10^{10}A_{{\rm s}})$ &  $0.1681 $  &  $0.3217 $  & $0.2209 $ & $0.0271$ &    $0.0240 $  &  $0.0259 $  & $0.0146 $\\
$H_{0}$                        &  $3.6902 $  &  $6.5433 $  & $4.0082 $ & $1.0341$ &    $0.5288 $  &  $0.6171 $  & $0.5282 $\\
$N_{\rm eff}$                  &  $1.7016 $  &  $3.3814 $  & $1.2486 $ & $--    $ &    $--     $  &  $--     $  & $--     $\\
$n_{{\rm s}}$                  &  $0.0201 $  &  $0.0727 $  & $0.0550 $ & $0.0046$ &    $0.0043 $  &  $0.0045 $  & $0.0039 $\\
$\Omega_{\nu}h^{2}$            &  $0.0044 $  &  $0.0048 $  & $0.0017 $ & $--    $ &    $ --    $  & $--      $  & $ --    $ \\\hline\hline
\end{tabular}
\end{center} 
\end{table*} 

\begin{figure}
\begin{center}
\includegraphics[width=8.8cm]{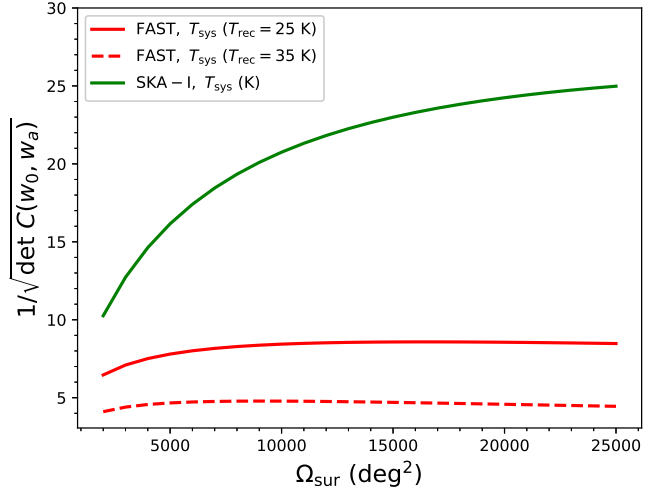} 
 \caption{Figure of merit (FoM): inverse square root of the determinant 
 of $w_{0}, w_{a}$ covariance matrix, $1/\sqrt{\text{det} \ C(w_0, w_a)}$
 versus survey area, $\Omega_{\rm sur}$ ($\text{deg}^{2}$) for FAST (red) 
 at system with receiver temperatures, $T_{\rm rec}$, of $25 \ \text{K}$ (solid line), $35 \ \text{K}$ 
 (dashed line) and for SKA-I (green) at system temperature, $T_{\rm sys}$ (K) given by Eq.~(\ref{SKAI_T_sys}).}
 \label{FOM}
 \end{center}
\end{figure}

In order to investigate the optimal survey volume, we consider FAST and SKA-I,
and explore the range of survey areas from $2,000 \ \text{deg}^{2}$ to $25,000 \ \text{deg}^{2}$. Considering $(w_{0}, w_{a})$ constraints, we find the optimal survey area is around $16,000 \ \text{deg}^{2}$ for FAST with $T_{\rm sys}$ corresponding to $T_{\rm rec} = 25 \ {\rm K}$ and $9,000 \ \text{deg}^{2}$ 
for a $T_{\rm sys}$ corresponding to $T_{\rm rec} = 35 \ {\rm K}$.
Results show that SKA-I can survey up to a maximum area of $25,000 \ \text{deg}^{2}$. 
This reality can be illustrated by the figure of merit (FoM) shown in Figure~\ref{FOM}.
Figure of merit is defined as \citep{FOM, generalized-FOM}
\begin{equation}
 \text{FoM} \propto [\sigma(w_0)\sigma(w_a)]^{-1}\propto 1/\sqrt{\text{det} \ C (w_0, w_a)}.
\end{equation}

We vary the survey area and see which $\Omega_{\rm sur}$ can maximize the FoM. As previously stated, we choose the survey area of $10,000 \ \text{deg}^{2}$ that was 
covered by the Sloan Digital Sky Survey (SDSS)~\citep{SDSS-DR11-DR12}. The choice has a benefit 
of being fairly moderate and is potentially suitable for comparative and cross-correlation studies
involving SDSS-like experiments, FAST and SKA-I. In addition, it is practical to choose this survey area for FAST and SKA-I comparisons because the marginal increase of FAST FoM is quite small if $\Omega_{\rm sur}>10,000$, so we will 
use $\Omega_{\rm sur}=10,000$ in our forecast. BINGO \citep{BINGO} can survey an approximate area 
of $2,500 \ \text{deg}^{2} - 3,000 \ \text{deg}^{2}$ as shown by \cite{Yi-Chao, FAST_IM}, 
but for this particular study, we use a survey area of $3,000 \ \text{deg}^{2}$ as was 
suggested by \cite{FAST_IM}.

Generally speaking, higher system temperature will result in higher noise spectra, which makes the constraints worse. This is well indicated by the figure of merit ( Fig.~\ref{FOM}),
as shown by the two FAST system temperatures, $T_{\rm sys}$ of $25 \ \text{K}$ and $35 \ \text{K}$.
Low values of $1/\sqrt{\text{det} \ C(w_0, w_a)}$ at high system temperature means that experimental 
performance decreases with an increase in system temperature. For this reason, it is likely 
that BINGO is mostly affected because of its high overall system temperature.

\begin{figure}
\begin{center}
 \includegraphics[width=8.8cm]{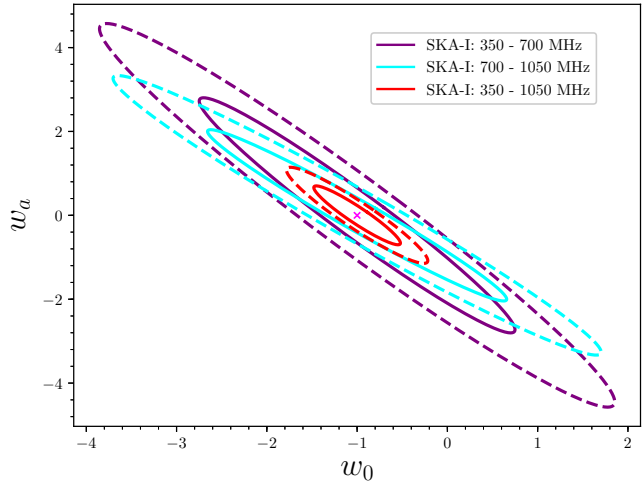} 
 \caption{$w_{0}$ versus $w_{a}$,  $1\sigma$ (solid line) and $2 \sigma$ 
 (dashed line) cosmological constraints for SKA-I split into lower frequency 
 band (purple), high frequency band 
 (cyan) and full range (red) 
 of SKA-I frequencies.}
 \label{splitted_w0_wa_FAST_BINGO_SKAI}
 \end{center}
\end{figure}

\begin{figure}
\begin{center}
\includegraphics[width=8.0cm]{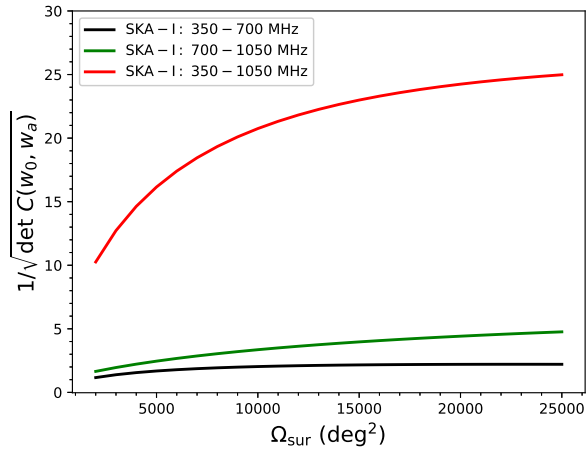} 
 \caption{Figure of merit (FoM): inverse square root of the determinant 
 of $w_{0}, w_{a}$ covariance matrix, $1/\sqrt{\text{det} \ C(w_0, w_a)}$
 versus survey area, $\Omega_{\rm sur}$ ($\text{deg}^{2}$) for various SKA-I frequency bands: 
 lower frequency band, $350 - 700 \ {\rm MHz}$ (black),  
 upper frequency band, $700 - 1050 \ {\rm MHz}$ (green), 
 and the full SKA-I frequency range, $350 - 1050 \ {\rm MHz}$ (red).}
 \label{SKAI_FOM}
 \end{center}
\end{figure}

There are several reasons why SKA-I performs better than both BINGO and FAST to constrain
the dark energy equation of state. One of the reasons is the SKA-I's wide range 
of frequency coverage. We split the SKA-I frequency range into
lower frequency band $350\,{\rm MHz}\sim 700\,{\rm MHz}$ and high frequency band 
$700\,{\rm MHz}\sim 1050\,{\rm MHz}$, and compare them with the full SKA-I frequency range ($350\,{\rm MHz}\sim 1050\,{\rm MHz}$). As shown in Figure~\ref{splitted_w0_wa_FAST_BINGO_SKAI}, 
the full SKA-I range of frequencies proportionately 
puts more stringent constraints on $w_0$ and $w_a$ than lower and upper frequency bands, 
and the FoM improves significantly, as shown in Figure~\ref{SKAI_FOM}. The reason is that the full frequency range of SKA-I includes the measurement of \hi power spectrum at larger range of redshift evolution, and 
also includes the information of cross-higher and lower frequency bands correlated signals. Therefore, it provides tighter constraints than higher and lower redshift bands.

\begin{figure}
\begin{center}
\includegraphics[width=8.0cm]{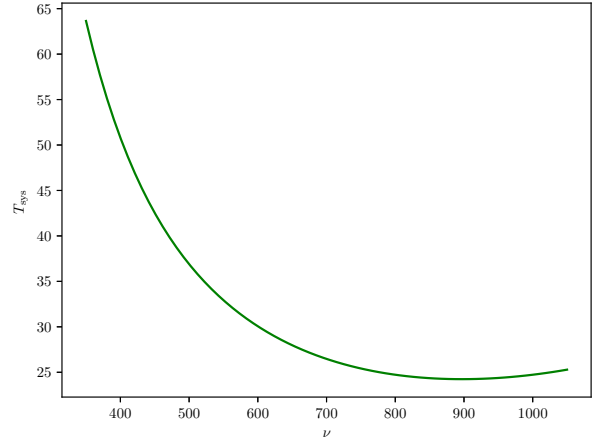} 
\caption{SKA-I variation of system temperature, $T_{\rm sys}$ against frequency, $\nu$.}
\label{SKAI_freq_vs_T_sys}
\end{center}
\end{figure}

Moreover, as we have previously accounted for BINGO, system temperature seems to be an important nuisance, which if not controlled, will severe constraints. The less stronger constraints for the 
SKA-I lower half of the frequency band compared to the upper band in Figure~\ref{splitted_w0_wa_FAST_BINGO_SKAI} (see also Figure~\ref{SKAI_FOM}) is suggestively due to high system temperatures at the corresponding 
frequencies.
The system temperature is somewhat a function of frequency, especially parts of $T_{\rm sys}$ (\ref{SKAI_T_sys})
that varies with it, i.e. Eqs.~(\ref{SKAI_T_rec}) and (\ref{SKAI_T_gal}). As a result, we see that system temperature, $T_{\rm sys}$ is more dominant at low frequencies than at high frequencies, see Figure~\ref{SKAI_freq_vs_T_sys}. 
This effect can as well be noted for FAST FOM at different system temperatures, Figure~\ref{FOM}.

The current SKA-I experimental design is to have 133 15-metre dishes and 64 13.5-metre MeerKAT dishes. We illustrate forecast with SKA-I by considering
the previous number of dishes, that's $190$ and then compare the constraint forecasts with the updated number of dishes, that's $133$. The reason for considering
the former number of dishes is to illustrate how the change in the number of dishes affects performance, and also to form a reference point for comparison with other previous literatures
which considered old SKA-I (SKAI-MID) experimental specifications. 

\begin{figure}
\begin{center}
 \includegraphics[width=8.8cm]{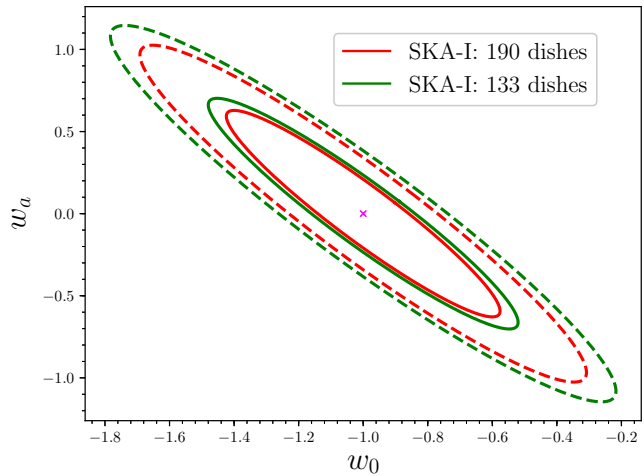} 
 \caption{$1\sigma$ (solid line) and $2\sigma$ (dashed line) comparisons of SKA-I (SKAI-MID Band I) constraints on the dark energy EoS by considering 
 early proposition of $190$ number of dishes (red),  and the updated (green)  $133$ number of dishes.}
 \label{compared_dishes_133_190}
 \end{center}
\end{figure}

We see from Figure~\ref{compared_dishes_133_190} that constraints do not strongly respond to the number of dishes. The most intrinsic property of large number of dishes is mapping LSS at very large angular distances/scales
by integrating \hi emission efficiently over large volumes slices of the sky. One would expect that large number of dishes would significantly improve cosmological constraints, but
if we compare constraints by assuming SKAI-MID $190$ and on the other hand $133$ number of single dishes, both in 
the autocorrelation mode, we notice not much significant difference.

The exact procedure of how to incorporate images from different frequency bands is unknown~\citep{SKAI_2018_Red_Book}, but in this forecast we assume that the SKA-I project is 
an integrated $190$ $15$-metre single-dishes in an autocorrelated mode for \hi intensity mapping.

\subsection{Constraints of other cosmological parameters}
\label{other_constraints}
\begin{figure*}
\centering
\includegraphics[width=18cm]{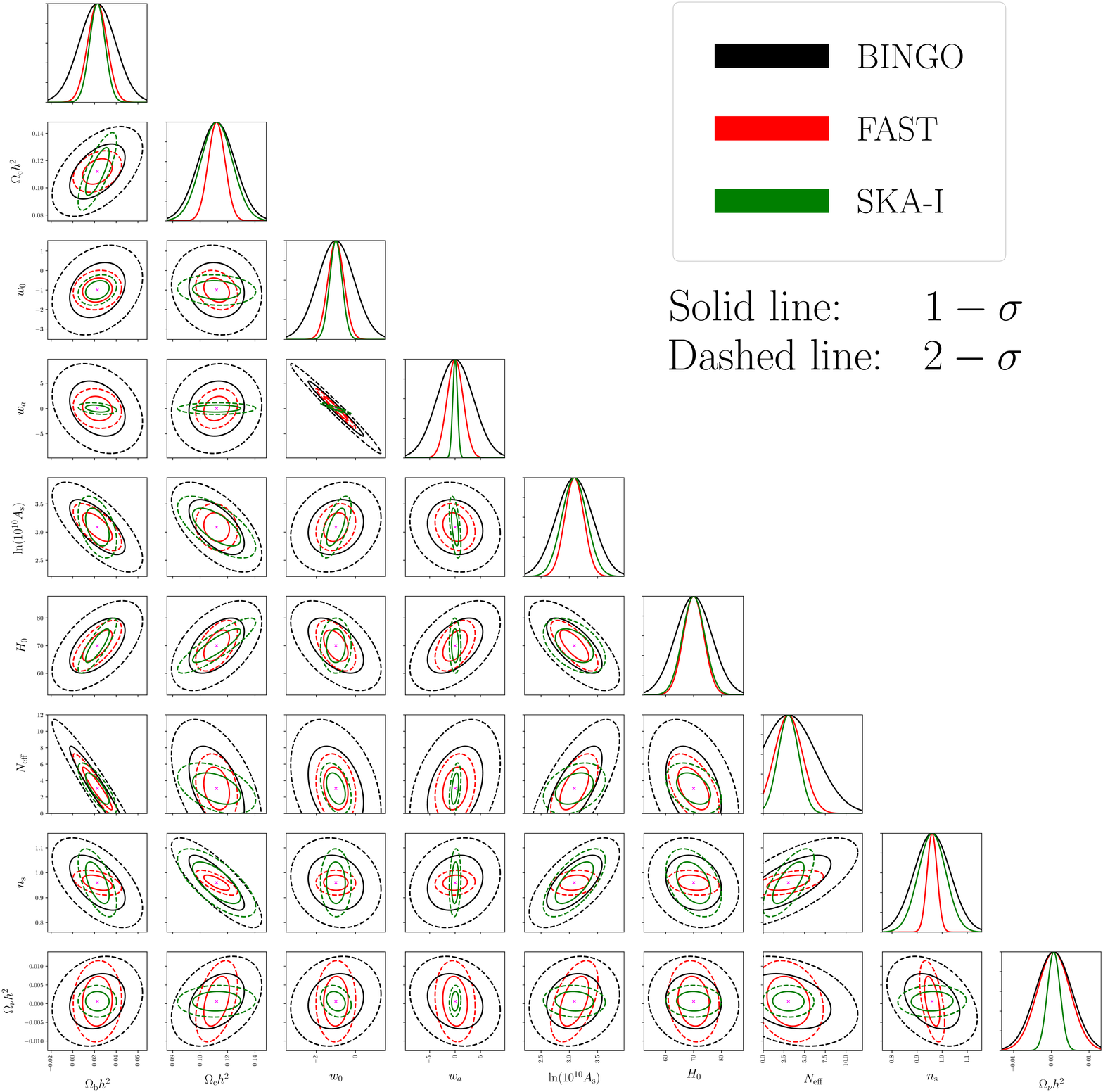}
\caption{Forecasts of cosmological constraints with FAST, BINGO 
and SKA-I future observations.} 
\label{FAST_BINGO_SKAI}
\end{figure*}

We present the results of our forecast for the $9$ cosmological parameters in Table \ref{cosmological_parms_description}
for the single dish experiments: FAST~\citep{Nan, FAST_IM}, BINGO~\citep{BINGO, BINGO2} and the SKA-I~\citep{SKAI, SKAI-2}. Figure~\ref{FAST_BINGO_SKAI} shows the constraints 
on various cosmological parameters. For the case of dark energy equation of state, 
we have seen that SKA-I will provide strongest constraints followed by FAST and then BINGO.

\begin{figure}
\centering
\includegraphics[width=8.8cm]{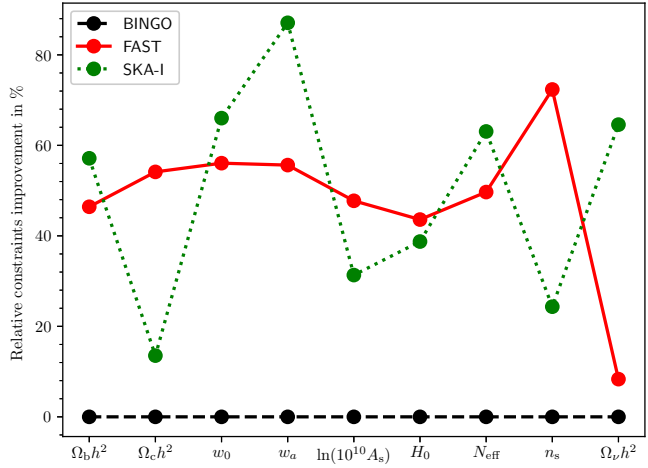}
\caption{The relative percentage improvement for FAST and SKA-I with respect to BINGO in constraining each of the $9$ cosmological parameters.} 
\label{relative_constraint_improvement}
\end{figure}

Considering all $9$ parameters, SKA-I and FAST are competitive in 
their abilities in constraining cosmological parameters. As shown in Figure~\ref{FAST_BINGO_SKAI},
FAST provides stronger constraints 
on $n_{\rm s}$ because its larger dish can provide more constraints on small-scales of \hi power spectra. Interestingly, the marginalized constraints on $n_{\rm s}$ for BINGO
and SKA-I do not show much significant difference. FAST will also impose stronger constraints on 
$\Omega_{\rm c}h^{2}$, $\text{ln}(10^{10}A_{\rm s})$ and $H_{0}$ than both BINGO and SKA-I, but slightly better constraint on $H_{0}$ than SKA-I. In comparison, SKA-I will strongly 
constrain $\Omega_{\nu}h^{2}$ in addition to $w_{a}$, while slightly better constraining $\Omega_{\rm b}h^{2}$ and $w_{0}$ parameters 
than FAST. Another observation is that SKA-I imposes slightly stronger bounds on both $N_{\rm eff}$ joint and marginalized constraints than FAST. 
The corresponding $\Omega_{\nu}h^{2}$, $N_{\rm eff}$ and $\Omega_{\rm b}h^{2}$  $1 \sigma$ errors 
for SKA-I, respectively, reduces by $61.36\%$,  $26.62\%$ and $20\%$ relative to FAST. Likewise, 
the corresponding $1 \sigma$ errors for the parameters where FAST performs better than SKA-I: $n_{\rm s}$, $\Omega_{\rm c}h^{2}$, $H_{0}$, $\text{ln}(10^{10}A_{\rm s})$ respectively, 
are reduced by $63.45\%$, $46.96\%$, $7.93\%$, $23.9\%$ relative to the corresponding SKA-I $1 \sigma$ errors. 
These reductions in the errors proportionately imply improvements 
in constraints as reflected by simulations. We depict in Figure~\ref{relative_constraint_improvement} the relative 
constraints improvement in percentage for all parameters and for the three simulated experiments.

The prospective better performance of FAST in constraining particular
parameters as we have seen, is due to its high angular resolving power. FAST 
has the largest dish diameter of the three telescopes which means its angular
resolution is higher than that of SKA-I and BINGO by a factor of $21$ and $7.5$ respectively, making it very capable to map 
signals at small angular scales. So, it is likely that FAST performance will be significant 
at small scales. On the other hand, Figure~\ref{overlap_frequency} indicates FAST is noise-dominated for $\ell>100$, 
since its signal-to-noise ratio (SNR) is less than unity,
while SKA-I does not attain SNR$<1$ until $\ell>150$. This suggests that SKA-I may better constrain cosmological parameters at some ranges of small angular scales due to its higher 
signal-to-noise ratio compared to FAST on those scales. Although SNR for SKA-I is greater than unity until $\ell>150$, from this point onwards,
SKA-I SNR decreases exponentially, while SNR for FAST decreases gently across the same range of scales. This draws another important clue that both FAST and SKA-I can relatively perform well
in constraining cosmological parameters sensitive to small angular scales. As we have previously pointed, the trade-off on whether FAST or SKA-I can perform better at small scales,
may not be determined by a single factor, but a number of factors, for example, the choice of parameterization.

For the case of dark energy constraints, SKA-I stringent constraints is due to its wide frequency coverage than FAST and BINGO. BINGO is probably underprivileged due to its high system/receiver noise temperature.

\begin{figure*}
\includegraphics[width=18cm]{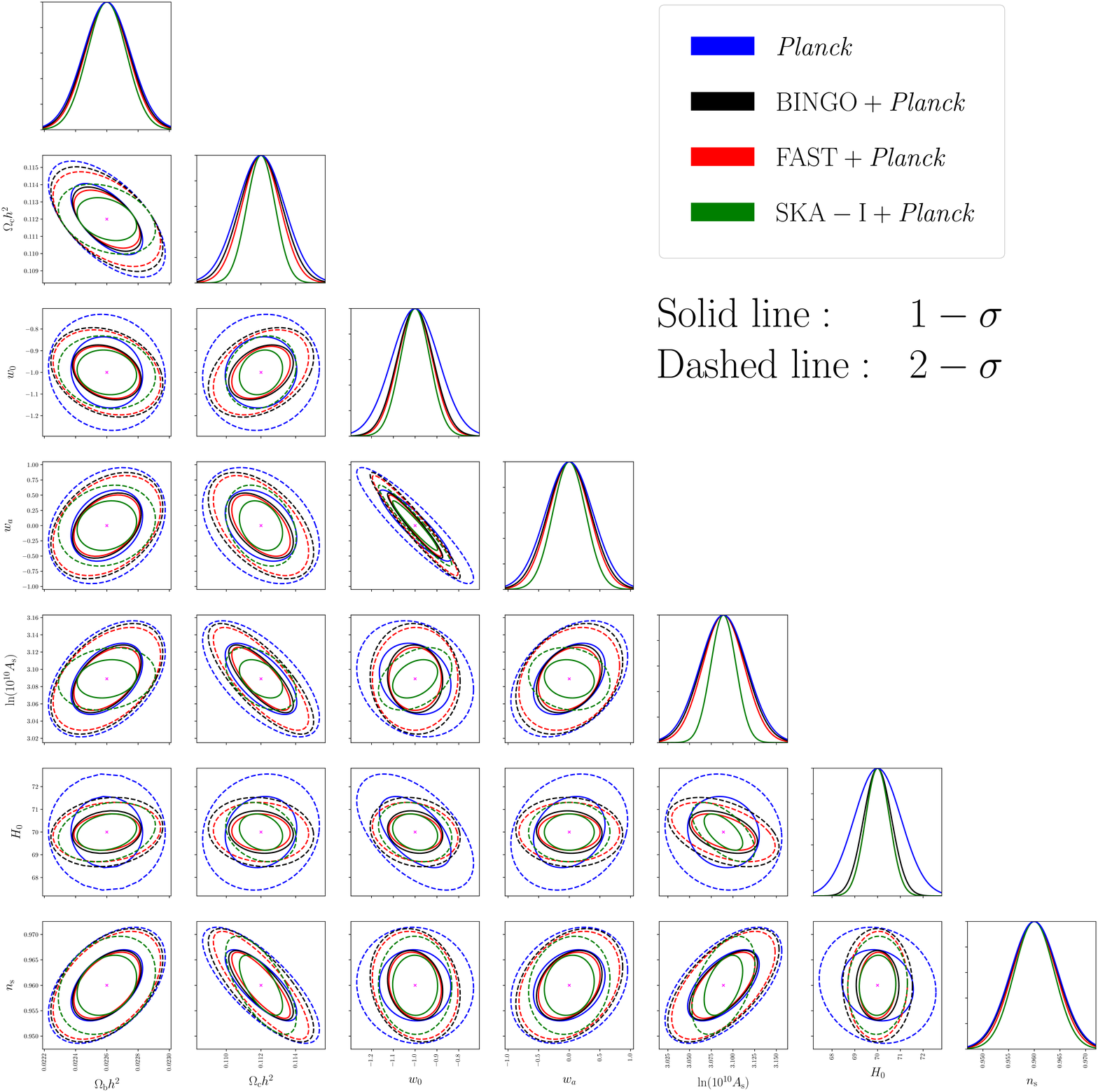}
\caption{Forecasts of joint cosmological constraints with each of the FAST, BINGO 
and SKA-I experiments plus {\it Planck} data, compared with {\it Planck} data constraints alone.} 
\label{FAST_BINGO_SKAI_plus_Planck}
\end{figure*}

Figure~\ref{FAST_BINGO_SKAI_plus_Planck} visualizes cosmological constraints 
for each of these experiments combined with \textit{Planck} by using Fisher 
matrix forecast. At this instance, constraints for all the three experiments 
improve significantly. As we have seen, SKA-I + \textit{Planck} continues to provide better constraints than
FAST + \textit{Planck} and BINGO + \textit{Planck} on dark energy equation of the state parameters $w_{0}$ and $w_{a}$. 
In contrast to the previous case involving FAST, BINGO and SKA-I experiments, 
SKA-I + \textit{Planck} constraints on $\text{ln}(10^{10}A_{\rm s})$ and $\Omega_{\rm c}h^{2}$ relative to FAST + 
\textit{Planck} and BINGO + \textit{Planck} experiments are now very significant. Similarly, neither FAST + \textit{Planck}
nor SKA-I + \textit{Planck} shows significant improvement in constraining $\Omega_{\rm b}h^{2}$ compared to BINGO + \textit{Planck}.
Therefore, SKA-I + \textit{Planck} imposes strong constraints on $\Omega_{\rm c}h^{2} \ {\rm and } \ \text{ln}(10^{10}A_{\rm s})$, in addition to $w_{0}, \ w_{a}$, as we have seen previously, relative to
FAST + \textit{Planck} and BINGO + \textit{Planck}.

More specifically, SKA-I + \textit{Planck} shows some significant improvement in constraining $\Omega_{\rm c}h^{2}, \ {\rm and } \ \text{ln}(10^{10}A_{\rm s})$ 
than FAST + \textit{Planck} and BINGO + \textit{Planck} by respectively, $27.27\%, \ 39.17\%$ and $33.33\%, \ 43.63\%$.
However, SKA-I + \textit{Planck} is respectively, slightly better in constraining $n_{\rm s}$ than FAST + \textit{Planck} and BINGO + \textit{Planck} by $9.3\%$ and $13.33\%$.

In the like manner, FAST + \textit{Planck} shows some significant improvement in constraining $\Omega_{\rm c}h^{2}$,  $w_{0}$, $w_{a}$, $\text{ln}(10^{10}A_{\rm s})$, 
$H_{0}$ and $n_{\rm s}$ than BINGO + \textit{Planck}  by $8.33\%$, $4.93\%$, $5.88\%$, $7.34\%$, $14.31\%$ and $4.44\%$, respectively. 

\begin{figure}
\centering
\includegraphics[width=8.8cm]{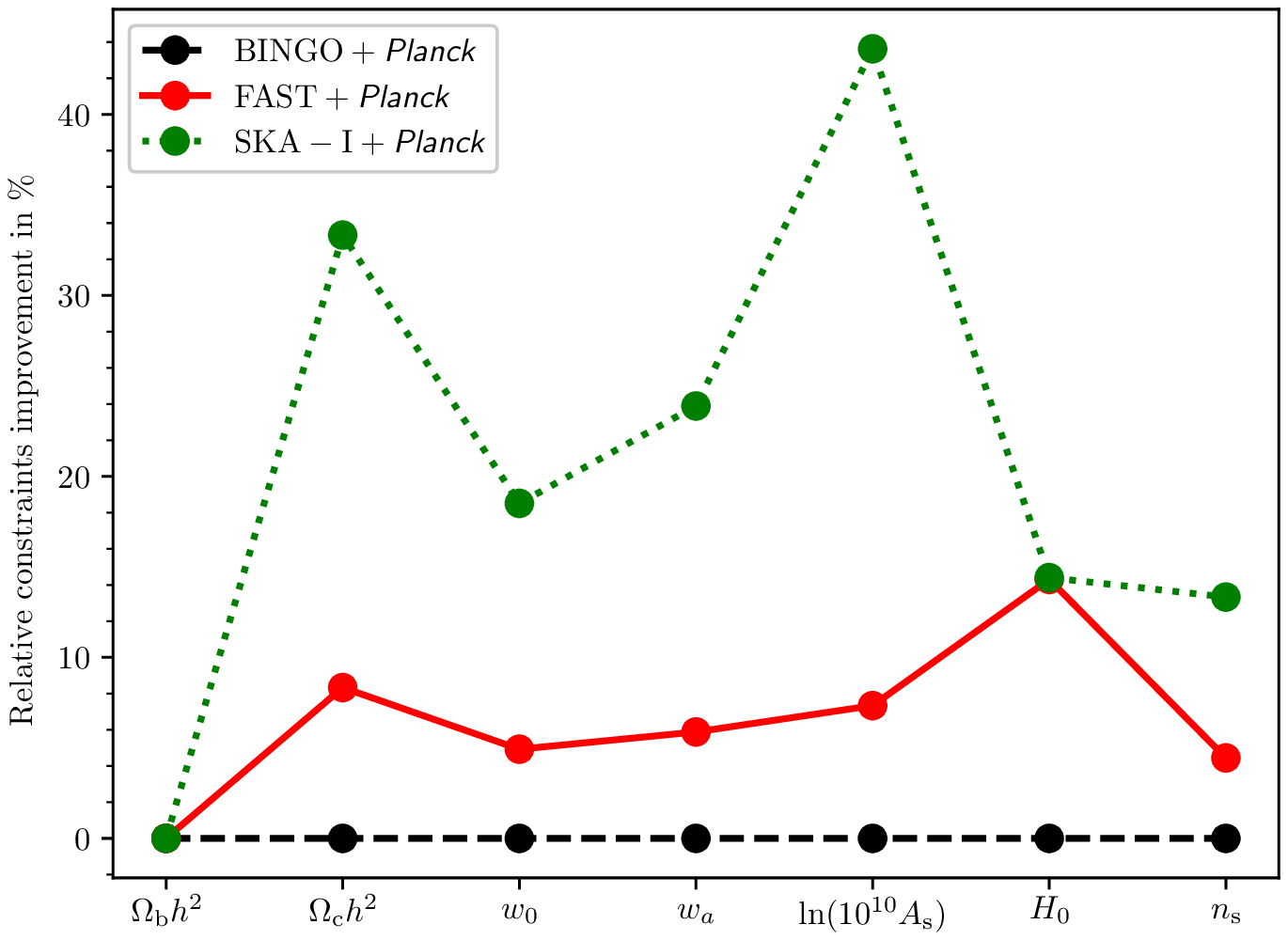}
\caption{The relative percentage improvement for FAST + \textit{Planck} and SKA-I + \textit{Planck} with respect to BINGO + \textit{Planck} in constraining each of the $7$ cosmological parameters
we have considered.} 
\label{BINGO_FAST_SKA_I_plus_Planck_rel_cons_improvement}
\end{figure}

Though there is significant performance improvement in constraining most of the cosmological
parameters for FAST + \textit{Planck} compared to the FAST alone, for SKA-I + \textit{Planck} compared to SKA-I alone and for BINGO + \textit{Planck} 
compared to BINGO alone, there is no improvement for FAST + \textit{Planck} in constraining $H_{0}$ relative to SKA-I + \textit{Planck}, see 
Figures~\ref{FAST_BINGO_SKAI_plus_Planck} and \ref{BINGO_FAST_SKA_I_plus_Planck_rel_cons_improvement}.

\begin{figure}
\centering
\includegraphics[width=8.8cm]{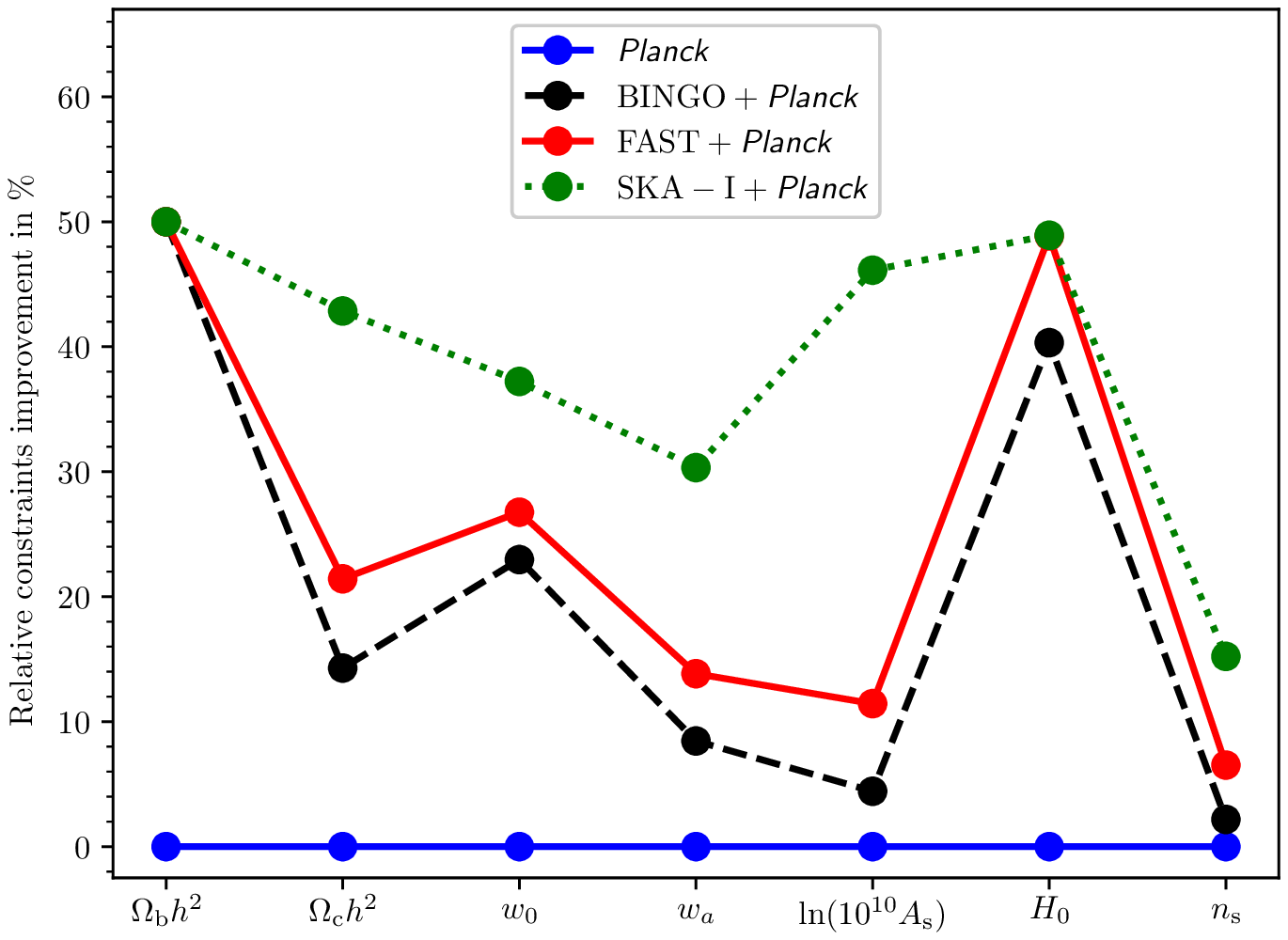}
\caption{The relative percentage improvement for BINGO + \textit{Planck}, FAST + \textit{Planck} and SKA-I + \textit{Planck} with respect to \textit{Planck} alone in constraining each of the $7$ cosmological parameters
we have considered.} 
\label{FBS_plus_Planck_rel_cons_improvement}
\end{figure}

As we have seen from Figures~\ref{FAST_BINGO_SKAI_plus_Planck} and \ref{BINGO_FAST_SKA_I_plus_Planck_rel_cons_improvement},
SKA-I + \textit{Planck}, followed by FAST + \textit{Planck}, are more competitive in 
constraining cosmological parameters than BINGO + \textit{Planck}. In any case, \textit{Planck} results
have very significant impact to constrain cosmological parameters when combining with each of the three experiments (Fig.~\ref{FBS_plus_Planck_rel_cons_improvement}).

\begin{figure}
\centering
\includegraphics[width=8.8cm]{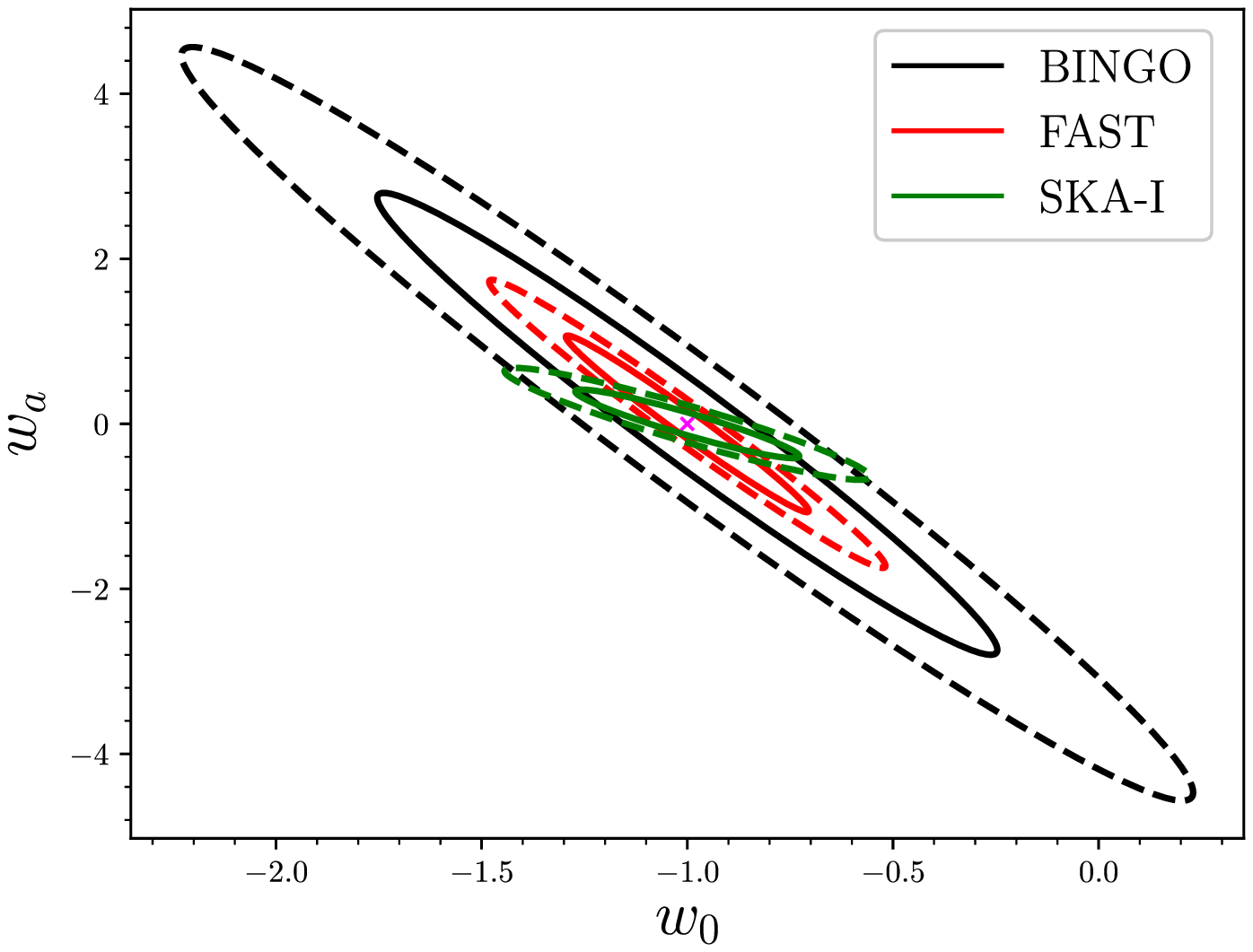}
\caption{Forecasts of cosmological constraints with FAST, BINGO and SKA-I with a frequency channelization, $\Delta \nu=1$ MHz.} 
\label{1MHz_freq_channel_FAST_BINGO_SKAI_23}
\end{figure}

In addition, we have tested using the dark energy EoS and find that, for BINGO, FAST and SKA-I \hi IM experiments, the choice of frequency channel width $\Delta \nu=1$ MHz can
significantly improve constraints for all the three \hi experiments than larger channel width (Fig.~\ref{1MHz_freq_channel_FAST_BINGO_SKAI_23}). This is because smaller band width can preserve the
redshift-space-distortion effect on radial direction
which makes it less ``Limber canceled'' than wider bandwidth~\citep{Hall13}. This is also illustrated in figure 6 of \citep{Xu17}.

\begin{figure}
\begin{center}
 \includegraphics[width=8.8cm]{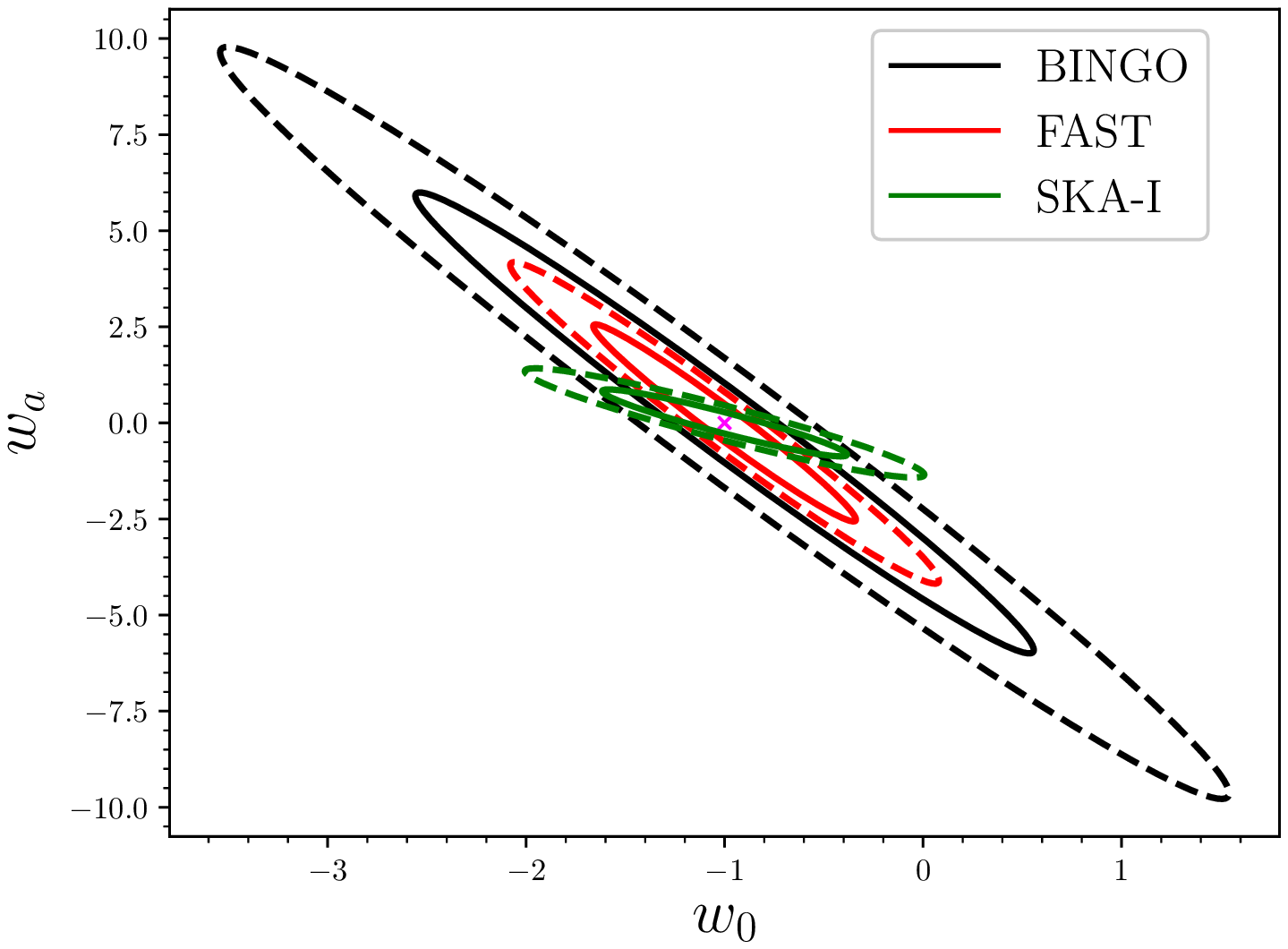} 
 \caption{$w_{0}$ versus $w_{a}$,  $1\sigma$ (solid line) and
 $2 \sigma$ (dashed line) cosmological constraints for 
 FAST (red), BINGO (black) and SKA-I (green) for minimum multipole moment, $\ell = 10$.}
 \label{min_ell_10_FAST_BINGO_SKAI}
 \end{center}
\end{figure}

\begin{figure}
\begin{center}
 \includegraphics[width=8.8cm]{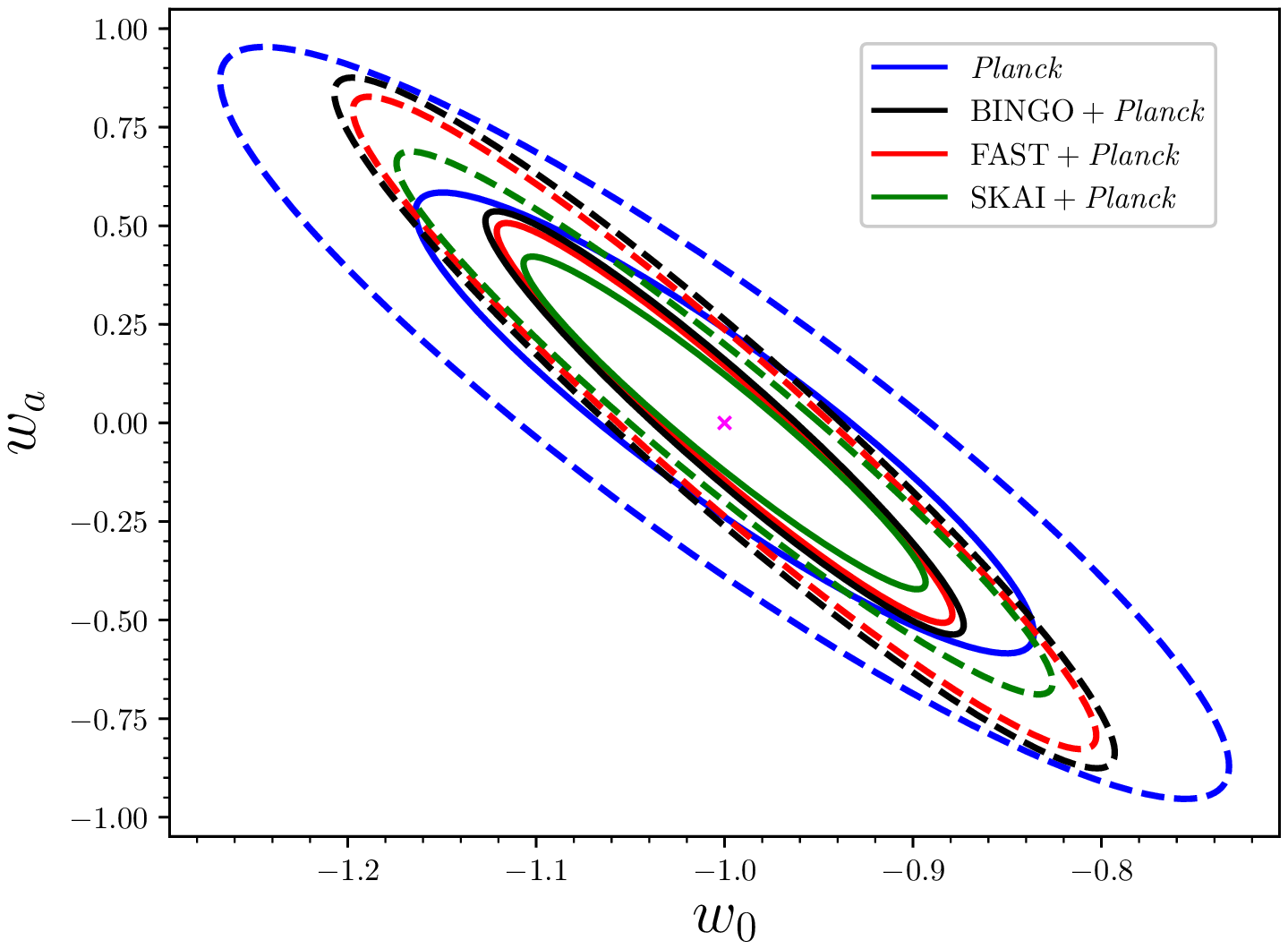} 
 \caption{$w_{0}$ versus $w_{a}$,  $1\sigma$ (solid line) and
 $2 \sigma$ (dashed line) cosmological constraints for {\it Planck} (blue)
 FAST + {\it Planck} (red), BINGO + {\it Planck} (black) and SKA-I + {\it Planck} (green) for minimum multipole moment, $\ell = 10$.}
 \label{min_ell_10_Planck_FBS_plus_Planck}
 \end{center}
\end{figure}

To strike a balance between limitation of foreground techniques to extract 
\hi signal at high angular scales and constraint prospects, as a case study, we simulate new dark energy EoS constraints 
by ignoring multipole moments at large angular scales, up to $\ell = 9$. The comparisons between Figures~\ref{w0_wa_FAST_BINGO_SKAI} \& \ref{w0_wa_FAST_BINGO_SKAI_plus_Planck} 
simulated by considering full multipole range of our interest, and Figures~\ref{min_ell_10_FAST_BINGO_SKAI} \& \ref{min_ell_10_Planck_FBS_plus_Planck}
where we apply  multipole moments cut-off, by considering  minimum $\ell = 10$, show a conflicting scenario between foreground effects
if we ignore small $\ell$'s and optimistic constraint forecast if we include them. Neglecting several values of $\ell$ corresponding to large angular scales weakens constraints,
and the $1 \sigma$ errors for marginalized $(w_{0}, w_{a})$ constraints, changes, respectively, for BINGO, FAST and SKA-I 
from $(0.9293, 3.5792)$, $(0.4083, 1.5878)$, $(0.3158,  0.4622)$ to $(1.0250, 3.9449)$, $(0.4355, 1.6864)$, $(0.4059, 0.5735)$;
and from $(0.0832, 0.3520)$, $(0.0791, 0.3313)$, $(0.0678, 0.2679)$ to $(0.0835, 0.3532)$, $(0.0795, 0.3336)$, $(0.0702, 0.2776)$ 
for BINGO + {\it Planck}, FAST + {\it Planck}, SKA-I + {\it Planck}, as summarized in Table \ref{min_ell_10_stds_FAST_BINGO_SKAI}. 

\begin{table*} 
\begin{center}
\caption{$1 \sigma$ errors for FAST, BINGO, SKA-I covariance 
matrices, and those obtained from covariance matrices resulting 
from combination of each of the FAST, BINGO and SKA-I experiment's 
Fisher matrix with {\it Planck} Fisher matrix, for minimum multipole moments, $\ell = 2$ and $\ell = 10$. Errors show that constraints resulting from discarding small values of $\ell$, i.e., $2\le \ell<10$, 
equivalent to large angular scales are weaker than those including small $\ell$'s. }
\label{min_ell_10_stds_FAST_BINGO_SKAI}
\begin{tabular}{@{}cccc|ccc} \hline\hline 
                                & FAST   &  BINGO  & SKA-I & FAST + {\it Planck} & BINGO + {\it Planck} & SKA-I + {\it Planck} \\\hline \hline
$\ell\ge 2$          & $ $ & $ $ &   $ $ & $ $ & $ $ & $ $ \\
$w_{0}$                          &   $0.4083   $  &  $0.9293   $  & $ 0.3158 $ &   $0.0791     $  &  $ 0.0832   $  & $ 0.0678 $ \\
$w_{a}$                          &   $1.5878   $  &  $3.5792   $  & $ 0.4622 $ &   $0.3313     $  &  $ 0.3520   $  & $ 0.2679 $\\
$\ell \ge 10$ & $ $ & $  $ & $ $ &   $ $ & $ $ & $   $\\
$w_{0}$                          &   $0.4355   $  &  $1.0250   $  & $ 0.4059 $ &   $0.0795     $  &  $ 0.0835   $  & $ 0.0702 $\\
$w_{a}$                  &           $1.6864   $  &  $3.9449   $  & $ 0.5735 $ &   $0.3336     $  &  $ 0.3532   $  & $ 0.2776 $\\
$ $            & $ $ & $ $ & $ $ & $   $ & $     $ & $   $ \\\hline\hline
\end{tabular}
\end{center} 
\end{table*} 

This is a clear illustration of how large angular scales which are more dominated by the foreground contaminations, may affect the cosmological constraint forecast analyses. We will however assume that the ongoing
progress in circumventing the foreground challenge and bias at large scales and other systematics at both large and small scales will be successful,
and hence allowing us to consider the maximum possible range of $\ell$'s, as we have done so under this study.

The subject of foreground in general, its dominion and removal 
challenge on certain angular scales has been discussed in \cite{fast_ICA, FASTICA2, Alonso_D_2015} and the references therein; whereby \cite{Planck_2018} presents a comparative study on the performance 
of a number of algorithms for diffuse component separation, just to name a few.

The current study is primarily focused on forecast of cosmological parameter constraints with \hi IM experiments. To this point, we postpone the in-depth discussion of the foregrounds challenge 
including addressing systematic limitations such as contaminations of single-dish observations by what is so called $1/f$ noise, a single instrumental systematic, due to frequency correlation, and the 
correlated gain fluctuations across the receiver bandpass~\citep{1_over_f_noise} to the next issue of this series.

\section{Comparison with previous forecasts of \hi IM}
\label{section_5}
This forecast aims to optimize future 21cm IM experiment potentials, 
by providing in-depth comparative objective study focusing on FAST, BINGO and SKA-I in autocorrelation mode - a collection of independent single-dish 
(rather than usual interferometry) telescopes. We use much cleaner and explicit  maximum likelihood and Fisher matrix tools to forecast
the behavior of these three telescopes by considering a wide range of sensitive experimental analyses aspects, laying
formalism that can be used to forecast varying sets of cosmological parameter constraints with a diverse range of 21cm IM experiments.
We notice that there are several previous studies that have made cosmological forecasts for \hi intensity mapping experiments, but our paper has the following distinctive features:

\begin{itemize} 
\item Extended the work by \cite{Late-time-cosmos} to consider different cosmological parameter set.
\cite{Late-time-cosmos} considered a set of standard $\Lambda$CDM  model: the Hubble parameter, $H_{0} = 100h \ {\rm km \ s^{-1} \ Mpc^{-1}}$, the cosmological constant, $\Omega_{\Lambda}$, the 
baryons density, $\Omega_{\rm b}h^2$, the linear amplitude of density fluctuations, $\sigma_{8}$,  the index of the power spectrum of primordial density fluctuations, $n_{\rm s}$, and the optical 
depth to last scattering, $\tau$. They extended the $\Lambda$CDM model with parameters $w_{0}, \ w_{a}, \  \Omega_{\rm K}$ and the growth index, $\gamma$. Here the cosmological constant, $\Omega_{\Lambda}$ 
and the curvature parameter, $\Omega_{\rm K}$ are related to the total matter density (Cold Dark Matter
+ baryons) by $\Omega_{\rm M} = 1 - \Omega_{\rm K} - \Omega_{\Lambda}$. In their forecast they used varying subsets of the considered parameters set to measure constraints. Their forecast approach included 
fixing fiducial values of some parameters, marginalizing over in the {\it Planck} priors Fisher matrix or over other parameters, not directly constraining some parameters by assuming their strong correlation 
with other parameters, such as in the case of {\it Planck} priors, where {\it Planck} measurements were combined with a particular experiment. We extended a subset of parameters considered in the 
aforementioned paper to form a new set, Table \ref{cosmological_parms_description}, and carried on Fisher matrix forecast, derived and treated under somewhat different approach. However, we expanded 
both cosmological parameters and \hi IM set of experiments compared to such papers as \cite{BINGO-IM} and \cite{BINGO, FAST_IM} to form a different forecast portfolio. Forecasting by considering various 
experimental designs and parameter sets is indispensable, since each set of cosmological parameters intertwined with a particular experimental design in principle, characterizes unique prediction results 
with an intention to harmoniously and comparatively contribute to address caveats and pinpoint prospects as we move towards a more precision and convergent cosmology.

\item Furthermore, our forecast incorporates more recent realistic and finalized development and design information, as these telescope constructions
have been undergoing major updates since the previous forecast results. These revisited experimental update set-ups, include the 
number of beams, dish diameter, frequency bandwidth coverage, survey area for FAST (see FAST included in the early study in \cite{Late-time-cosmos}); and number of dishes for SKA-I, updated confirmed information 
about its precursor, MeerKAT and the new approach for modeling system temperatures.

For example, previous forecasts with SKA-I considered $190$ dishes, while we make comparison, for illustration purpose using the case of dark energy EoS (see Fig.~\ref{compared_dishes_133_190}) in terms 
of SKA-I performance by considering old and updated number of dishes, we 
use the recently accepted and confirmed dishes for SKA-I from \cite{SKAI_2018_Red_Book} for comparative study with BINGO and FAST. 

However, a number of previous forecasts were limited by the information made publicly available during that time. These updates are crucial, 
because the whole essence of forecast is to enable the \hi IM experiments to optimize their performances by considering each aspect and every single detail of their experimental designs 
and specifications to find out how each experiment is sensitive to various variables. 

\item We use a reasonably narrow and computationally effective frequency channelization with a bandwidth 
channel of $10 \ {\rm MHz}$ each as contrasted to previous forecasts, such as \cite{Late-time-cosmos} which considered $60 \ {\rm MHz}$ for all experiments. 
Our consideration accounts for the role of narrower channel bandwidths, as expected for the modern radio receivers \citep{Late-time-cosmos} in tightening the constraints. 

\item We forecast for even more narrower frequency channel width of $1 \ {\rm MHz}$. This choice is close to the expected channelization of the future real \hi IM surveys \citep{Nan, FAST_IM}.

\item In order to break degeneracies and improve precision of cosmological constraints, we include $\textit{Planck}$ 2015 CMB priors measurements that have been rigorously tested and improved, they include,
CMB lensing reconstruction; TT, TE, EE \textit{Planck} Cosmic Microwave Background (CMB) (\cite{WMAP}) power spectra; where TT represents temperature power spectrum, TE is 
temperature-polarization cross-spectrum, and EE is polarization power spectrum; and high $\ell$ CMB measurements. 
This was not objectively considered by even the forecasts such as those which tried to include as many experiments as possible.

\item We have provided more extensive  quantification of cosmological constraints forecast in regard to these representative telescopes of our choice focusing on \hi IM surveys. Other related works such as 
\cite{Villaescusa-Navarro} have studied the BAOs measurements through a single-dish \hi IM observations in the post-reionization epoch in the light of SKAI-MID. \cite{all-sky-interferometry, coaxing-21cm} 
have alternatively addressed forecast of cosmological  constraints, \hi power spectrum estimation and measurement analyses of wide-field transient telescopes such as CHIME by an approach of what they call m-mode 
formalism. Furthermore, \cite{Pourtsidou_2017} forecasted \hi evolution with redshift and a select subset of model-independent cosmological parameters focusing on the performance of the SKA and its precursor 
MeerKAT \hi IM surveys \citep{MeerKAT_IM}, proposing their cross-correlation with optical galaxy surveys. Under different setting, constraints on 
the dark energy parameters by cross-correlating/combining SKA-like \hi IM and LSST-like surveys have been performed by \cite{Pourtsidou_2015}. 

The essence of cross-correlating \hi IM or 21cm maps in general
with galaxy redshift surveys, is that contaminants, such as foregrounds, various noises and systematics between the maps from two types of surveys are largely expected to be uncorrelated in frequency; this is 
in contrast to \hi signal which is correlated in respective frequency bands. As a result, cross-correlation will statistically boost the abundance and the amplitude of \hi signal, but also statistically cancel 
out relevant foregrounds and systematics, thus increasing \hi signal detection, and consequently improve constraints on the estimated values of the cosmological and astrophysical parameters. Constraints of 
$\Omega_{\rm HI}b_{\rm HI}r$ have direct link to the future IM surveys capabilities and the prospect science outputs, and these surveys heavily depend on the qualitative and quantitative measurements 
(such as shape and amplitude) of \hi signal. Cross-correlation will thus aggregate more \hi signal information than any individual experiments, yielding robust and precise cosmological measurements 
\citep{MeerKAT_IM}. \cite{Pourtsidou_2017}, for example, has reported improvement to about a factor $\sim 3$ by considering Stage IV spectroscopic galaxy survey (similar to {\it EUCLID}) and MeerKAT 
with an overlap area of $500\ {\rm deg}^{2}$ in constraining the amplitude of the quantity $\Omega_{\rm HI}b_{\rm HI}r$, where $r$ is a correlation coefficient that accounts for possible 
stochasticity in the galaxy and \hi tracers \citep{MeerKAT_IM}. Similarly, with an overlap area of $4000 \ {\rm deg}^{2}$, cross-correlation between MeerKAT and Stage III photometric optical galaxy 
survey measured/constrained $\Omega_{\rm HI}b_{\rm HI}r$ at $\sim 5$ percent level across a wide range of redshifts compared to the autocorrelated MeerKAT constraints. According to them, such 
improvements were better than autocorrelation results they could achieve. The fact that the cross-correlated power spectrum will be less sensitive to contaminations, can be used to identify systematics 
in 21cm maps \citep{Wolz_2016, Pourtsidou_2017, Carucci_2017}.
Cross-correlation could be less susceptible to systematic contaminants \citep{Pourtsidou_2015}, hence foregrounds and systematics are expected to be highly suppressed, respectively, making their removal
and control much easier.

\cite{Furlanetto_2007} has laid down several advantages of cross-correlation, two of them are: firstly, the signal-to-noise ratio resulting from cross-correlating 21cm experiments and galaxy redshift 
surveys exceeds that of the individual 21cm power spectrum by a factor of few, further asserting that, this may allow probing of smaller spatial scales and possibly more efficient detection of inhomogeneous 
reionization. Secondly, the approach highly reduces the required level of foreground cleaning for the 21cm signal/maps. \hi IM and galaxy redshift survey cross-correlation approach to suppress foregrounds and 
systematics has also been motivated, explored and echoed using simulations by a number of other authors, some of them include \cite{Wolz_2016, Carucci_2017, Cunnington_2019}. This observation is also supported 
by \cite{Pen_2009, 2010Natur.466..463C, 2013MNRAS.434L..46S, GBT, 2017arXiv171000424A} who achieved the detection of \hi by cross-correlating the \hi IM and optical galaxy redshift surveys. Synergized 
cross-correlation between these two types of surveys has mutual benefits, that make them complement each other in alleviating survey-specific systematic effects and boost \hi signal detection.

Other forecasts include CMB bounds on $f_{\rm NL}$ by combining information from SKA Phase I and {\it Euclid}/LSST-like photometric galaxy surveys 
using multi-tracer, contrasting with respective single-tracer measurements \citep{Fronseca_2015}; and an extension of this approach for \hi IM 
with MeerKAT and photometric galaxy survey to constrain $f_{\rm NL}$ and a number of other parameters \citep{Fonseca_2017}. %considered combination and individual exp 

\item Although combination of different subsets of cosmological parameters and experimental designs largely characterize the future telescope performances, this study has singled out those features
intrinsic to the particular experiment and are likely to determine their performance reliability, consistency and stability in benchmarking with other similar surveys.
 \end{itemize} 

In this paper, we have, therefore, intentionally addressed forecasts of cosmological constraints 
for the three \hi IM experiments under consideration, while including issues previously not given 
significant attention, updating the forecasts to suit the upgrades undergone by the considered 
telescopes and individually and simultaneously comparatively asses the three telescope 
performances while laying down a basis for any other cosmological constraints forecast 
with \hi IM experiments, as we prepare for real survey take-off with these next generation instruments. 
A great deal of useful information we aggregate through our researches play a complementary role in 
building a scientific body of knowledge that can be maximally deployed to continually study the Universe.

\section{Conclusion}
\label{section_6}

We have conducted forecasts for cosmological constraints (Figs.~\ref{w0_wa_FAST_BINGO_SKAI}, 
\ref{w0_wa_FAST_BINGO_SKAI_plus_Planck}, \ref{FAST_BINGO_SKAI}, \ref{FAST_BINGO_SKAI_plus_Planck}) 
for a set of $9$ cosmological parameters (Table \ref{cosmological_parms_description}) and compared performance for three different
proposed future survey projects, FAST, BINGO and SKA-I. Our results, with a prescribed 
choice of experimental parameter set (Table \ref{experimental_parameters}) show that 
FAST experiment will have better performance compared to BINGO, particularly, 
in constraining dark energy equation of state. In overall, SKA-I will put more 
stringent constraints for the dark energy equation of state than FAST and BINGO. 
We notice that, there is a trade-off between SKA-I and FAST in constraining cosmological 
parameters, with each experiment being more superior in constraining a particular set 
of parameters.

We point out that narrower frequency bandwidth such as $1 \ \text{MHz}$ 
(see Fig.~\ref{1MHz_freq_channel_FAST_BINGO_SKAI_23}) greatly improves constraints because 
the redshift-space-distortion effect suffers less cancellation if 
frequency band becomes narrower~\citep{Hall13,Xu17}. But this requires more
computer resources in terms of memory (RAM) for intermediate storage and 
speed for reasonable computational time. This challenge can however be 
addressed by advancing computing  resources and modeling strategy. 
We postulate that, high frequency resolution needs one to take into 
account correlated noise residues at $i$-th and $j$-th frequency bins
which would become noticeable due to many frequency channels being 
correlated, otherwise noise residues would be significant to be 
ignored and in some way impact the results. However, real 
instrumentation will use much narrower frequency bandwidth 
which would facilitate radio frequency interference excision \citep{Nan}.

We conclude that for a single-dish approach, BINGO, FAST and SKA-I 
will progressively provide stronger constraints on dark energy 
equation of the state and other cosmological parameters. 
The constraints can be further improved by combining with CMB 
experiment such as {\it Planck} data.

We performed \hi IM Fisher matrix forecast for BINGO, FAST and SKA-I radio telescopes, and extended 
similar comparative analysis for each of the three experiments' data, combined with \textit{Planck} 
chains, \citep{Planck-collaboration} which have considerably tighten the cosmological constraints. 
This substantial and objective comparative analysis of simulated forecast results provides a benchmark 
on the relative expected performances of BINGO, FAST and SKA-I experiments under 
relatively similar settings in constraining an extended number of cosmological 
parameters in Table \ref{cosmological_parms_description}. FAST, BINGO, SKA-I and many other telescopes are 
suitable for \hi IM, and some will even do a wide range of sciences \citep{Nan} than others. 
Our aim is not to show the superiority or inferiority of these experiments against each other, but to
illustrate a global picture on their relative prospects. Our results can however, signal for 
adjustment, revision of specification configurations, or for further calibration where there 
is a possibility in order to rectify and optimize capabilities so that these telescopes can 
fulfil their promise.

This paper sets an important mark for our series of works to study 
IM surveys with HI. Future proceedings will feature applicability
and quantification of this novel but a promising approach by 
developing IM pipeline to simulate 
sky maps for various sky emissions, addressing and testing different 
foreground cleaning methods, investigating and quantifying various
calibration issues. These realistic issues include  bandpass calibration, 
systematics and other uncertainty measurements, studying and 
developing solid knowledge of polarization purity, and measuring BAO wiggles 
from \hi power spectrum and consequently developing more 
stringent constraints on dark energy, dark matter and other cosmological parameters. 

\section{Acknowledgements}
E.Y. acknowledges the DAAD (German Academic Exchange Service) scholarship and the financial support from The African Institute for Mathematical Sciences, University of KwaZulu-Natal, and 
The Dar Es Salaam University College of Education, 
Tanzania. Y.C.L. and Y.Z.M. acknowledge support from the National Research Foundation of South Africa (Grant No. 105925 and 110984).

%% you can type \apj for ApJ, \aap for A&A, \apss for Ap&SS, etc. Please consult
%% the macro chjaa.cls. You can also find them in aasguide.tex (AASTeX for ApJ, AJ, PASP)
%% Please follow the format of ChJAA's reference list
\bibliographystyle{raa}
\bibliography{RAA_Manuscript}
%\label{lastpage}
\end{document}